\documentclass[
 reprint,
 amsmath,amssymb,
 aps,prx
]{revtex4-2}

\usepackage[english]{babel}
\usepackage{amsmath}
\usepackage{bm}
\usepackage{amsfonts}
\usepackage[cal=cm]{mathalpha}
\usepackage{nicematrix}
\usepackage{graphicx}
\usepackage{tcolorbox}
\usepackage[colorlinks=true, allcolors=blue]{hyperref}
\usepackage{amsthm}
\usepackage{algorithm}
\usepackage{algpseudocode}
\usepackage{epstopdf}
\usepackage{booktabs}
\usepackage{tabularx}
\usepackage{capt-of}
\usepackage{longtable}
\usepackage{array}
\usepackage{ragged2e}
\usepackage{float}
\usepackage{placeins}

\begin{document}

\title{Complexity Reveals the Microscopic Origins of Macroscopic Dynamics}

\author{Haoyang Qian$^{1}$}
\author{Beata Casiday$^{2}$}
\author{Gabriel Hood$^{3}$}
\author{Malbor Asllani$^{1}$}
\email{masllani@fsu.edu}
\affiliation{$^{1}$Department of Mathematics, Florida State University, Tallahassee, FL, USA}

\affiliation{$^{2}$Yale University, New Haven, CT, USA}

\affiliation{$^{3}$Georgia Institute of Technology, Atlanta, GA, USA}

\begin{abstract}
Real complex systems often exhibit collective transitions emerging from interactions across many components. Classical stability theory describes such transitions in spectral space, where dynamics is organized by spatially extended global eigenmodes whose collective nature obscures direct association with individual physical components. Here, we show that structural disorder in empirical random networks can fundamentally alter this picture. These properties induce spectral localization, causing Laplacian modes to concentrate on small subsets of nodes and producing a mode--node correspondence in which collective dynamics becomes governed predominantly by the local behavior of a dominant node together with their effective coupling to the surrounding network. As a consequence, stability properties can be interpreted directly in node space rather than purely in spectral space. Exploiting this principle, we develop a node-resolved framework that predicts transition onsets, identifies the nodes responsible for emergent collective behavior, and restores interpretability in systems where classical modal theories fail. In heterogeneous reaction networks, the same mechanism gives rise to exotic collective states where different subsets of nodes develop distinct dynamical behaviors beyond those associated with homogeneous assumptions. Our results show that complex network structures naturally generate spectral localization, revealing the microscopic drivers of macroscopic dynamics.
\end{abstract}

\maketitle

\section{Introduction}
\label{sec:intro}

Complex systems are inherently heterogeneous. In biological, ecological, technological and social networks, components rarely obey identical dynamics: nodes differ in parameters, timescales, interaction strengths, and often even in the governing equations themselves \cite{newman2010networks,sugitani2021synchronizing,sun2009master,acharyya2012synchronization,nazerian2023synchronization}. At the same time, interactions are highly irregular, producing sparse and disordered network architectures with broad structural variability. Yet much of the theory of collective dynamics has traditionally relied on homogeneous settings, where identical units and symmetric couplings allow the dynamics to be decomposed into independent spectral modes. These approaches have revealed fundamental mechanisms of synchronization, spreading and pattern formation \cite{MSF,arenas_synchronization_2008, nakao_turing_2010,asllani_theory_2014,diego2018key, van2023emergence}, but they become fundamentally limited to apply in heterogeneous systems, where the dynamics can no longer be cleanly separated into independent collective modes \cite{panahi2021group,acharyya2015synchronization,ricci2012onset,lu2010cluster,belykh2003persistent}. As a consequence, stability and collective organization remain naturally described in spectral space rather than in terms of identifiable physical components of the system.

In continuous media and networked systems alike, emergence is driven by instabilities and phase transitions whose dynamics is naturally organized in spectral space \cite{cross_pattern_2009,MSF,nakao_turing_2010,asllani_theory_2014}. Near continuous transitions, unstable modes determine not only the tipping points but also the spatial organization of emerging nonlinear patterns. Understanding which structural elements disproportionately govern emergence is therefore a central challenge across nonlinear dynamics, from transient amplification and emergent stability to nonlinear basin resilience \cite{menck2013basin, asllani2018structure, meena2023emergent}. Yet in conventional settings these unstable modes remain naturally described in spectral coordinates rather than in terms of identifiable physical components.

At the same time, substantial effort has been devoted to identifying the nodes that disproportionately influence collective behavior in complex systems. In epidemic spreading, synchronization, and network control, structural properties such as degree, centrality, community organization, or controllability have revealed how specific subsets of nodes can shape global dynamics \cite{pastor2001epidemic_2,liu2011controllability}. These approaches have provided important insight into influential or functionally relevant components, but they typically remain statistical or structural in nature and are not directly derived from the unstable modes organizing the dynamics itself.

Recent studies have shown that structural disorder in real networks generates strong spectral localization~\cite{mcgraw,nakao_loc,pastor2016distinct,metz2021localization}. Rather than remaining spatially extended, Laplacian eigenvectors can become strongly concentrated around restricted regions of a network, as shown for the trade network in Fig.~\ref{fig:Fig1}(a). Such localization naturally arises in large graphs and emerges naturally from randomness, degree variability, and structural disorder~\footnote{See Supplementary Material for extended empirical evidence and statistical analysis across a large dataset of real-world networks.}. As a consequence, the spectral basis progressively loses its globally extended character and individual modes acquire a spatial identity dominated by restricted subsets of nodes. In the strongly localized regime, Laplacian eigenvectors approach an almost node-resolved representation, behaving nearly as canonical basis vectors associated with specific physical components of the network.

Here we show that this localization structure naturally gives rise to a mode--node correspondence, whereby unstable modes can be mapped back onto identifiable physical nodes governing the collective dynamics. Building on this principle, we develop a localization reduction framework that replaces the full heterogeneous Jacobian by a set of node-resolved reduced problems. In this way, stability analysis can be interpreted directly in node space rather than purely in spectral space. The resulting framework predicts transition onsets, identifies the dominant nodes and effective couplings underlying collective instabilities, and establishes, to our knowledge, a first direct connection between spectral instabilities and node-level dynamical attribution in complex network systems. Figure~\ref{fig:Fig1} illustrates how this correspondence relates collective instabilities directly to identifiable subsets of nodes.

Beyond predicting emergence thresholds, the reduction reveals how spectral localization organizes nonlinear collective behavior beyond the criticality threshold. Across heterogeneous dynamical systems on networks, different localized instability structures give rise to exotic patterns such as cluster synchronization, amplitude chimeras, amplitude--phase chimeras, and localized oscillon-like states \cite{belykh2008cluster,sorrentino2016complete,Siebert,sethia_amplitude-mediated_2013,chimera,Abrams_Strogatz,Panaggio_2015,zakharova_chimera_2020,symm_break,umbanhowar_localized_1996,Vanag_Epstein}. Because localized modes can be independently destabilized through their dominant nodes, collective dynamics progressively disentangles into differentiated node-level behaviors, providing both mechanistic understanding and a route toward distributed control. Our results suggest that structural disorder in complex networks can restore physical interpretability to collective dynamics through spectral localization.

%%%%%%%%%%%%%%%%%%%%%%%%%%%%%%%%%
%%%%%%%%%%%%%%%%%%%%%%%%%%%%%%%%%

\section{Results}

We consider a broad class of networked dynamical systems composed of $N$ interacting subsystems, each described by an $M$-dimensional state vector $\mathbf{x}_i(t)$. Such systems arise naturally across biological, ecological, social, technological, and information settings, where large numbers of heterogeneous units interact through complex and often random patterns of connectivity~\cite{newman2010networks}. The collective dynamics is shaped both by the intrinsic evolution of each subsystem and by its interactions with other subsystems mediated by the network structure, and can be written in the general form
$
\dot{\mathbf{x}}_i
= \mathbf{f}_i(\mathbf{x}_i)
+ \boldsymbol{\sigma}_i \sum_{j=1}^N \mathcal{A}_{ij}\,\mathbf{G}(\mathbf{x}_i,\mathbf{x}_j)\,\,\,\forall i
\label{eq:main}
$
where $i$ labels the nodes of the network. Here, $\mathbf{f}_i$ is a vector-valued function describing the intrinsic dynamics of subsystem $i$, while the coupling function $\mathbf{G}$ is a vector-valued interaction term encoding how the state of node $j$ influences node $i$. The adjacency matrix $\mathcal{A}$ encodes the topology of pairwise interactions among subsystems. The matrix $\boldsymbol{\sigma}$ collects the coupling strengths across the system and is block diagonal, with each block $\boldsymbol{\sigma}_i$ acting on the $M$ components of node $i$ and controlling how strongly that node interacts with its neighbors. Importantly, heterogeneity is not limited to variations in parameter values: the functions $\mathbf{f}_i$ may differ across nodes not only in their parameters, but also in their nonlinearities and even in their functional form, thereby allowing fundamentally different dynamical models to coexist within the same network. For clarity, we develop the present framework in the undirected setting, which has long served as the standard paradigm for studying collective behavior in complex systems. At the same time, real-world networks are fundamentally directed and can exhibit hierarchical organization~\cite{johnson2017looplessness,asllani2018structure,Duan_Motter}. Nevertheless, recent work has shown that directed and empirical networks can exhibit strong spectral localization and localization-driven collective dynamics~\cite{muolo_persistence_2024}, suggesting that the underlying mode--node correspondence persists beyond symmetric interactions and motivating future extensions to directed networks.

\begin{figure*}[t]
\centering
\includegraphics[width=\textwidth]{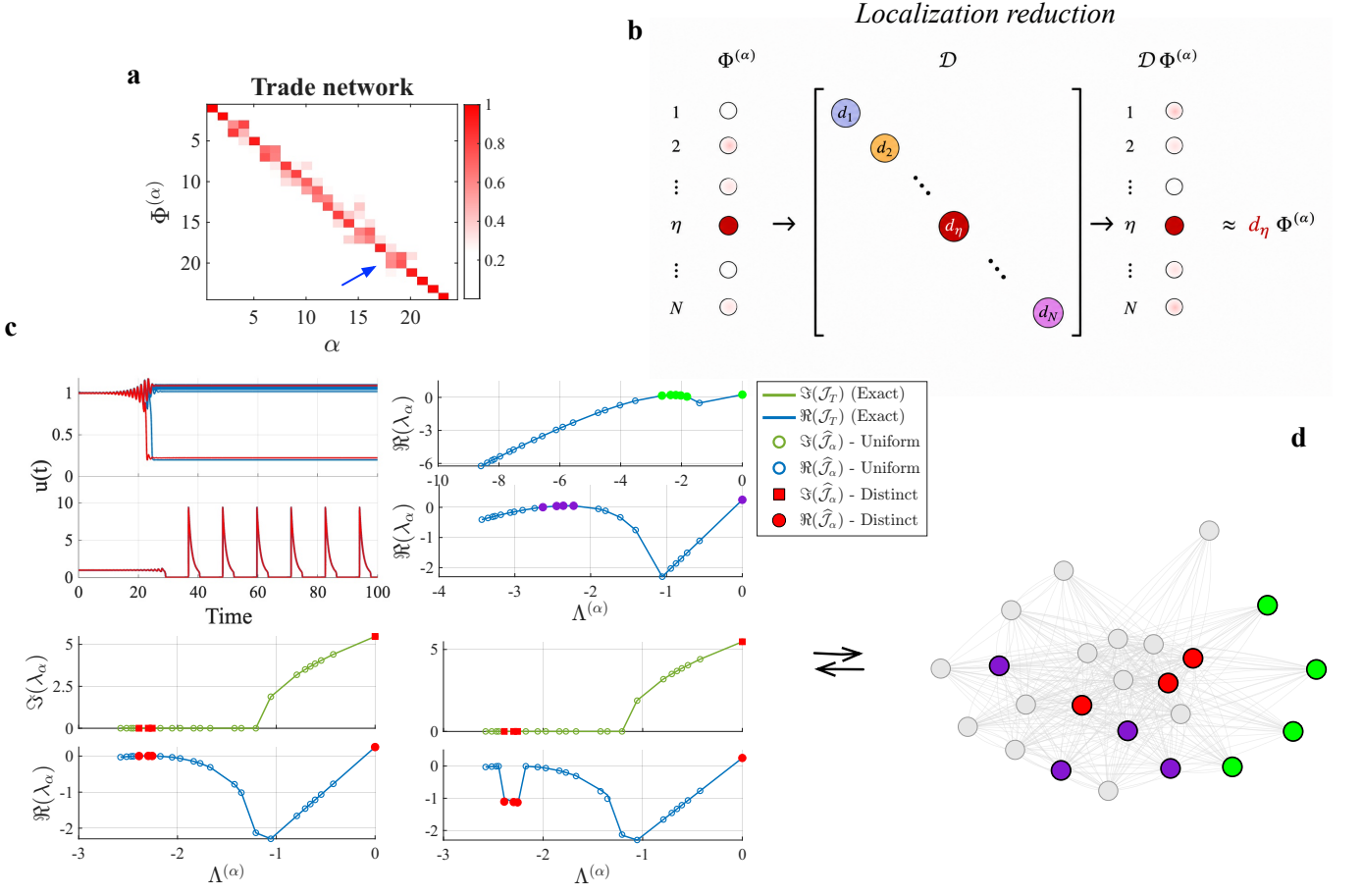}
\caption{\textbf{Localization-based mode--node mapping reveals node-driven dynamics.}
(\textbf{a}) Heatmap of the Laplacian eigenvector matrix in absolute value for the International Trade Network of Manufactured Goods \cite{DeNooy2018}, showing strong spectral localization. The blue arrow highlights an overlap region where multiple localized modes are associated with the same dominant node.
(\textbf{b}) Schematic illustration of the localization reduction: when a Laplacian eigenvector $\Phi^{(\alpha)}$ localizes around node $\eta$, the action of a diagonal heterogeneous matrix $\mathcal{D}$ is approximated by
$\mathcal{D}\Phi^{(\alpha)} \approx d_{\eta}\Phi^{(\alpha)}$, establishing an approximate correspondence between localized spectral modes and dominant physical nodes.
(\textbf{c}) Master Stability spectra and corresponding time series for the Brusselator model on the trade network. 
Green, purple, and red correspond to $(D_v,\sigma)=(2.821,0.12)$, $(2.83,0.16)$, and $(2.86,0.4)$, respectively. 
In the red case, three localized modes become unstable and synchronization is lost. 
(\textbf{d}) Network visualization highlighting the dominant nodes associated with the unstable localized modes shown in panel (\textbf{c}). 
Targeted modification of these nodes restores stability and synchronization in the corresponding time series, illustrating the approximate equivalence between localized instability modes and node-level dynamical drivers.}
\label{fig:Fig1}
\end{figure*}

\subsection{Heterogeneous Master Stability Function}

To study the emergence of heterogeneous collective behavior, it is natural to formulate the dynamics relative to uniform reference configurations whose destabilization drives network-induced transitions. Let $k_i=\sum_{j=1}^N \mathcal{A}_{ij}$ denote the degree of node $i$, and define the diagonal degree matrix $\mathcal{K}=\mathrm{diag}(k_1,\dots,k_N)$. Introducing the graph Laplacian $\mathcal{L}=\mathcal{A}-\mathcal{K}$, so that $\mathcal{A}=\mathcal{K}+\mathcal{L}$, the dynamics can be written as
\begin{equation}
\dot{\mathbf{x}}_i
=
\mathbf{F}_i(\mathbf{x}_i)
+
\boldsymbol{\sigma}_i
\sum_{j=1}^N
\mathcal{L}_{ij}\,\mathbf{G}(\mathbf{x}_i,\mathbf{x}_j),
\quad \forall i,
\label{eq:laplacian_form}
\end{equation}
where $\mathbf{F}_i(\mathbf{x}_i)
=
\mathbf{f}_i(\mathbf{x}_i)
+
\boldsymbol{\sigma}_i k_i \mathbf{G}(\mathbf{x}_i,\mathbf{x}_i)$ denotes the intrinsic dynamics renormalized by the self-interaction induced by connectivity. In this representation the Laplacian term vanishes under uniform configurations, ensuring invariance of the homogeneous state. This follows from the spectral property of the Laplacian: because it has zero row sum, it admits the uniform vector $\mathbf{1}$ as an eigenvector associated with the zero eigenvalue. Instabilities can therefore arise only through transversal eigenmodes associated with nonzero eigenvalues. This spectral structure isolates the network-induced mechanisms responsible for the emergence of spatial or dynamical heterogeneity. Notably, even when the original system is parametrically identical across nodes and coupling strengths are uniform, the effective dynamics $\mathbf{F}_i$ inherits heterogeneity through the degree $k_i$, showing that degree variability directly contributes to collective behavior.

To assess stability, we linearize Eq.~\eqref{eq:laplacian_form} around a reference solution by setting $\mathbf{x}_i=\mathbf{x}_i^\ast+\boldsymbol{\xi}_i$ and retaining terms to first order. Whenever the node dynamics admit a common homogeneous reference state $\mathbf{x}_i^\ast=\mathbf{x}^\ast$, the Laplacian structure ensures invariance of the homogeneous manifold.  Such uniform states, ranging from synchronized equilibria to homogeneous oscillatory solutions, act as organizing centers for collective transitions. We therefore restrict attention to transversal perturbations, which break uniformity and govern the onset of heterogeneous patterns.
Denoting by $D\mathbf{F}_i$ the Jacobian of the effective node dynamics and by $D_1\mathbf{G}$ and $D_2\mathbf{G}$ the Jacobians of the coupling function with respect to its first and second arguments, respectively, the linearized dynamics simplifies to
\begin{equation}
\dot{\boldsymbol{\xi}}_i
=
D\mathbf{F}_i(\mathbf{x}^\ast)\,\boldsymbol{\xi}_i
+
\boldsymbol{\tilde{\sigma}}_i
\sum_{j=1}^N
\mathcal{L}_{ij}\,\boldsymbol{\xi}_j,
\label{eq:lin_Deff}
\end{equation}
where the contribution proportional to $D_1\mathbf{G}$ vanishes because it is independent of the summation index and therefore annihilated by the Laplacian operator. Also, the effective coupling matrix is defined as \(\boldsymbol{\tilde{\sigma}}_i=\boldsymbol{\sigma}_iD_2\mathbf{G}(\mathbf{x}^\ast,\mathbf{x}^\ast),\) which collects the node-dependent coupling strength and the derivative of the interaction function into a single prefactor. The resulting expression separates intrinsic stability from Laplacian-mediated coupling effects and forms the basis for a spectral characterization of collective instabilities. 

To exploit the Laplacian structure without rewriting the dynamics node-by-node, 
we work in a species-stacked representation. For each species 
$k=1,\ldots,M$, let $\boldsymbol{\xi}^{(k)}(t)\in\mathbb{R}^N$ collect the 
$k$-th component across the network. The linearized dynamics \eqref{eq:lin_Deff} can now be written 
componentwise in species space as
\[
\dot{\boldsymbol{\xi}}^{(k)}(t)
=
\sum_{\ell=1}^{M}
\mathcal{D}^{(k\ell)}\,\boldsymbol{\xi}^{(\ell)}(t)
+
\sum_{\ell=1}^{M}
\mathcal{S}^{(k\ell)}\,\mathcal{L}\,\boldsymbol{\xi}^{(\ell)}(t),
\quad \forall k,
\]
where the $N\times N$ diagonal matrices
\begin{align*}
\mathcal{D}^{(k\ell)}
&:=
\mathrm{diag}\!\big(
[D\mathbf{F}_1(\mathbf{x}^\ast)]_{k\ell},\ldots,
[D\mathbf{F}_N(\mathbf{x}^\ast)]_{k\ell}
\big),
\\[.5em]
\mathcal{S}^{(k\ell)}
&:=
\mathrm{diag}\!\big(
[\boldsymbol{\tilde{\sigma}}_1]_{k\ell},\ldots,
[\boldsymbol{\tilde{\sigma}}_N]_{k\ell}
\big),
\end{align*}
collect across all nodes the $(k,\ell)$ entries of the local Jacobians
$D\mathbf{F}_i(\mathbf{x}^\ast)$ and effective coupling matrices
$\boldsymbol{\tilde{\sigma}}_i$. In other words, for each species pair
$(k,\ell)$, the diagonal entries of $\mathcal{D}^{(k\ell)}$ and
$\mathcal{S}^{(k\ell)}$ describe how the local dynamics and effective coupling coefficients vary across nodes.
Stacking the species-level perturbations into the vector
\[
\boldsymbol{\Xi}
=
\big(
(\boldsymbol{\xi}^{(1)})^\top,\ldots,(\boldsymbol{\xi}^{(M)})^\top
\big)^\top
\in\mathbb{R}^{MN},
\]
the linearized dynamics can be written compactly as
\begin{equation}
\dot{\boldsymbol{\Xi}} = \mathcal{J}_T\,\boldsymbol{\Xi},
\label{eq:tot_lin}
\end{equation}
where \(\mathcal{J}_T\in\mathbb{R}^{MN\times MN}\) is the full Jacobian of the
networked system in the species-stacked representation. In this ordering,
\(\mathcal{J}_T\) takes the natural \(M\times M\) block form
\[
\mathcal{J}_T =
\begin{pmatrix}
\mathcal{D}^{(11)}+\mathcal{S}^{(11)}\mathcal{L} & \cdots & \mathcal{D}^{(1M)}+\mathcal{S}^{(1M)}\mathcal{L} \\
\vdots & \ddots & \vdots \\
\mathcal{D}^{(M1)}+\mathcal{S}^{(M1)}\mathcal{L} & \cdots & \mathcal{D}^{(MM)}+\mathcal{S}^{(MM)}\mathcal{L}
\end{pmatrix}.
\]

We order the Laplacian eigenpairs $\{(\Lambda^{(\alpha)},\Phi^{(\alpha)})\}_{\alpha=1}^N$ so that $\Lambda^{(1)}=0$ 
corresponds to the homogeneous mode. We now expand each species-level perturbation in the Laplacian eigenbasis,
\[
\boldsymbol{\xi}^{(k)}(t)
=
\sum_{\alpha=1}^{N}
c_\alpha^{(k)}(t)\,\boldsymbol{\Phi}^{(\alpha)},
\quad \forall k.
\]
Substituting this expansion into the species-stacked equation and using
the eigenvalue–eigenvector relationship introduced above, we obtain
\begin{align}
\sum_{\alpha=1}^{N}
\dot c_\alpha^{(k)}(t)\,\boldsymbol{\Phi}^{(\alpha)}
&=
\sum_{\ell=1}^{M}\sum_{\alpha=1}^{N}
c_\alpha^{(\ell)}(t)\,
\mathcal{D}^{(k\ell)}\,\boldsymbol{\Phi}^{(\alpha)}
\label{eq:MST_stuck}
\\
&\quad +
\sum_{\ell=1}^{M}\sum_{\alpha=1}^{N}
\Lambda^{(\alpha)}\,
c_\alpha^{(\ell)}(t)\,
\mathcal{S}^{(k\ell)}\,\boldsymbol{\Phi}^{(\alpha)},
\quad \forall k.
\nonumber
\end{align}

This is precisely where the classical MSF decoupling breaks: although
$\mathcal{L}$ acts diagonally on $\boldsymbol{\Phi}^{(\alpha)}$, the
diagonal operators $\mathcal{D}^{(k\ell)}$ and $\mathcal{S}^{(k\ell)}$
are not multiples of the identity matrix when $D\mathbf{F}_i$ and
$\boldsymbol{\tilde{\sigma}}_i$ depend on $i$. Consequently,
$\mathcal{D}^{(k\ell)}\boldsymbol{\Phi}^{(\alpha)}$ (and
$\mathcal{S}^{(k\ell)}\boldsymbol{\Phi}^{(\alpha)}$) is not generally
proportional to $\boldsymbol{\Phi}^{(\alpha)}$, and projecting onto the
Laplacian eigenbasis produces coupling between different mode indices
$\alpha$.

\subsection{From mode decoupling to node-resolved dynamics: Localization reduction method}

As introduced earlier, and as widely documented in random complex networks, 
Laplacian eigenvectors are often strongly localized.
Rather than being uniformly distributed across the network, many nontrivial
modes concentrate their weight on a small subset of nodes.
This phenomenon becomes particularly pronounced in networks with structural heterogeneity,
degree variability, or mesoscopic organization. In such settings,
a transversal eigenmode typically exhibits a dominant component supported
on a single node (or a very small cluster of nodes), while its amplitude
elsewhere remains negligible.

Motivated by this spectral property, we define for each non-uniform mode
$\alpha \ge 2$ its \emph{host node} as
\begin{equation}
\eta(\alpha)
=
\arg\max_{1 \le i \le N}
\left|
\Phi^{(\alpha)}_{i}
\right|.
\label{eq:arg_local}
\end{equation}
Such localization implies that
\[
\left|
\Phi^{(\alpha)}_{i}
\right|
\ll
\left|
\Phi^{(\alpha)}_{\eta(\alpha)}
\right|
\quad
\text{for most } i \neq \eta(\alpha),
\]
so that the eigenvector $\boldsymbol{\Phi}^{(\alpha)}$
is effectively supported on node $\eta(\alpha)$ to leading order.
This separation of amplitudes provides the key simplification exploited below. 

Recall from Eq.~\eqref{eq:MST_stuck} that heterogeneity enters through the
diagonal matrices $\mathcal{D}^{(k\ell)}$ and $\mathcal{S}^{(k\ell)}$,
whose diagonal entries collect the node-dependent Jacobians and effective
coupling coefficients. Since these operators are diagonal in node space,
their action on a localized eigenvector is dominated by the entry
corresponding to the host node:
\begin{align*}
\mathcal{D}^{(k\ell)} \boldsymbol{\Phi}^{(\alpha)}
&\approx
[DF_{\eta(\alpha)}(\mathbf{x}^*)]_{k\ell}
\,\boldsymbol{\Phi}^{(\alpha)},
\\[.5em]
\mathcal{S}^{(k\ell)} \boldsymbol{\Phi}^{(\alpha)}
&\approx
[\tilde{\sigma}_{\eta(\alpha)}]_{k\ell}
\,\boldsymbol{\Phi}^{(\alpha)}.
\end{align*}
In the relations above, 
$DF_{\eta(\alpha)}(\mathbf{x}^*)$ denotes the $M \times M$ Jacobian 
matrix of the local reaction dynamics at the host node $\eta(\alpha)$, 
evaluated at the common equilibrium $\mathbf{x}^*$. 
Likewise, $\tilde{\sigma}_{\eta(\alpha)}$ is the corresponding 
$M \times M$ effective coupling matrix at that same node. 
Thus, the node-space diagonal operators 
$\mathcal{D}^{(k\ell)}$ and $\mathcal{S}^{(k\ell)}$ 
reduce, under localization, to the species-space matrices 
$DF_{\eta(\alpha)}(\mathbf{x}^*)$ and 
$\tilde{\sigma}_{\eta(\alpha)}$ 
associated with the node on which the Laplacian eigenvector 
$\boldsymbol{\Phi}^{(\alpha)}$ is concentrated. 
In this sense, each localized mode inherits the linearized dynamics of its host node.
This localization mechanism is illustrated schematically in Fig.~\ref{fig:Fig1}(b): when a transversal Laplacian mode concentrates around a dominant host node, the action of heterogeneous diagonal operators becomes effectively controlled by the local dynamics of that node.

Substituting these approximations into Eq.~\eqref{eq:MST_stuck}
eliminates the coupling between different mode indices $\alpha$.
Each localized mode then evolves independently as
\[
\dot{c}^{(k)}_{\alpha}
=
\sum_{\ell=1}^{M}
[DF_{\eta(\alpha)}(\mathbf{x}^*)]_{k\ell}
\, c^{(\ell)}_{\alpha}
+
\Lambda^{(\alpha)}
\sum_{\ell=1}^{M}
[\tilde{\sigma}_{\eta(\alpha)}]_{k\ell}
\, c^{(\ell)}_{\alpha}.
\]
Collecting terms, the modal amplitude vector\\
\(
\mathbf{c}_{\alpha}
=
\big(
c^{(1)}_{\alpha},
\dots,
c^{(M)}_{\alpha}
\big)^\top
\)
satisfies
\[
\dot{\mathbf{c}}_{\alpha}
=
\Big(
DF_{\eta(\alpha)}(\mathbf{x}^*)
+
\Lambda^{(\alpha)}\,
\tilde{\sigma}_{\eta(\alpha)}
\Big)
\mathbf{c}_{\alpha}.
\]
This defines a mode- and node-dependent Jacobian
\begin{equation}
\widehat{\mathcal{J}}_{\alpha}
=
DF_{\eta(\alpha)}(\mathbf{x}^*)
+
\Lambda^{(\alpha)}\,
\tilde{\sigma}_{\eta(\alpha)}.
\label{eq:J_alpha}
\end{equation}
The corresponding localized Master Stability Function, defined through the maximal growth rate \footnote{For stationary homogeneous states, the growth exponents correspond to the eigenvalues of the local Jacobian, while for homogeneous oscillatory solutions they correspond to the Lyapunov exponents of the synchronous solution.} associated with the localized transversal mode, is therefore
\begin{equation}
\Psi\left(\Lambda^{(\alpha)},\eta(\alpha)\right)
=
\max_{\lambda \in \operatorname{spec}(\widehat{\mathcal{J}}_{\alpha})}
\Re(\lambda),
\label{eq:localized_MSF}
\end{equation}
and instability occurs whenever $\Psi(\Lambda^{(\alpha)}, \eta(\alpha))>0$.
Unlike the classical MSF, the localized formulation depends explicitly on the
host node $\eta(\alpha)$ through the node-specific matrices
$DF_{\eta(\alpha)}(\mathbf{x}^*)$ and
$\tilde{\sigma}_{\eta(\alpha)}$.
As a result, instability can be attributed directly to the node on which
the corresponding Laplacian mode is localized.

%%%%%%%%%%%%%%%%%%%%%%%%%%%%%%%%%%%%%%%%%%%%%%%%%%%%%%%%%%%%%%%%%%%%%
%%%%%%%%%%%%%%%%%%%%%%%%%%%%%%%%%%%%%%%%%%%%%%%%%%%%%%%%%%%%%%%%%%%%%

\subsection{Node-resolved collective instabilities}

To illustrate how spectral localization identifies the nodes driving nonlinear collective dynamics, we consider a heterogeneous two-species reaction--diffusion system (two dynamical variables per node, $M=2$) admitting either stable homogeneous states or self-sustained limit-cycle oscillations. The dynamics takes the Laplacian reaction--diffusion form
\begin{equation}\label{eq:bruss_RD}
\begin{aligned}
\dot{u}_i &= f(u_i,v_i,a_i,b_i,c_i,d_i)
+\sigma_i^{(u)}\sum_j \mathcal{L}_{ij}u_j,\\
\dot{v}_i &= g(u_i,v_i,a_i,b_i,c_i,d_i)
+D_i^{(v)}\sum_j \mathcal{L}_{ij}v_j.
\end{aligned}
\end{equation}
The simplicity of the Laplacian coupling allows the localization mechanism to be isolated transparently, while more general coupling architectures and mixed dynamical models are discussed in the Supplementary Material (SM). Specifically, we study the networked Brusselator model defined in Methods~\ref{sec:bruss}, where heterogeneity is introduced through node-dependent reaction coefficients and diffusion rates. The nonlinear functions \(f\) and \(g\) preserve the same Brusselator structure across the network, while heterogeneity enters through the node-dependent parameters \(a_i,b_i,c_i,d_i\) and diffusion coefficients.

Starting from a baseline parameter set $(a_0,b_0,c_0,d_0)$ admitting a homogeneous fixed point $(u^*,v^*)$, we introduce node-dependent rescalings of the reaction kinetics,
\[
a_i = h_{1,i} a_0, \quad d_i = h_{1,i} d_0,
\qquad
b_i = h_{2,i} b_0, \quad c_i = h_{2,i} c_0 ,
\]
where $h_{1,i},h_{2,i}>0$. This paired rescaling preserves the same equilibrium across all nodes while allowing local kinetic timescales to vary. Consequently, the emergence of instability and spatial organization is governed by diffusion, network structure, and spectral localization rather than by trivial shifts of the local equilibria.

These dynamical implications are shown in Fig.~\ref{fig:Fig1}(c--d), where the localized Master Stability spectra identify the critical nodes across the three dynamical regimes shown in different colors. To compare the exact and reduced spectra, eigenvalues are optimally paired in the complex plane using the procedure described in Methods~\ref{sec:pairing}. In the present setting, the homogeneous mode associated with the zero Laplacian eigenvalue $\Lambda^{(1)}=0$ remains unstable with nonzero imaginary part, generating synchronized limit-cycle oscillations supported by the uniform eigenvector. Synchronization is therefore determined by the stability of the transversal localized modes associated with nonzero Laplacian eigenvalues. Rather than remaining distributed across extended spectral modes, the instabilities become identified with physical nodes through the localization mapping of the corresponding Laplacian eigenvectors. The host nodes associated with the unstable localized modes are highlighted in Fig.~\ref{fig:Fig1}(d) using the same color notation, linking unstable spectral modes directly to node-level dynamical drivers. In the red regime, multiple localized transversal modes become unstable simultaneously, leading to the loss of synchronization. Targeted modification of the local diffusion parameters at those nodes shifts the unstable modes back into the stable region, thereby restoring synchronization of the network dynamics while preserving the underlying synchronized oscillatory state. This illustrates how spectral localization not only predicts instability onset, but also provides a direct route toward node-level interpretation and distributed control of collective behavior.

%%%%%%%%%%%%%%%%%%%%%%%%%%%%%%%%%%
%%%%%%%%%%%%%%%%%%%%%%%%%%%%%%%%%%

%%%%%%%%%%%%%%%%%%%%%%%%%%%%%%%%%%%%%%%%%%%%%%%%%%%%%%%%%%%%%%%%%%%%%
%%%%%%%%%%%%%%%%%%%%%%%%%%%%%%%%%%%%%%%%%%%%%%%%%%%%%%%%%%%%%%%%%%%%%

\subsection{Data validation}
\label{sec:data}

\begin{figure*}[t!]
    \includegraphics[width=\textwidth]{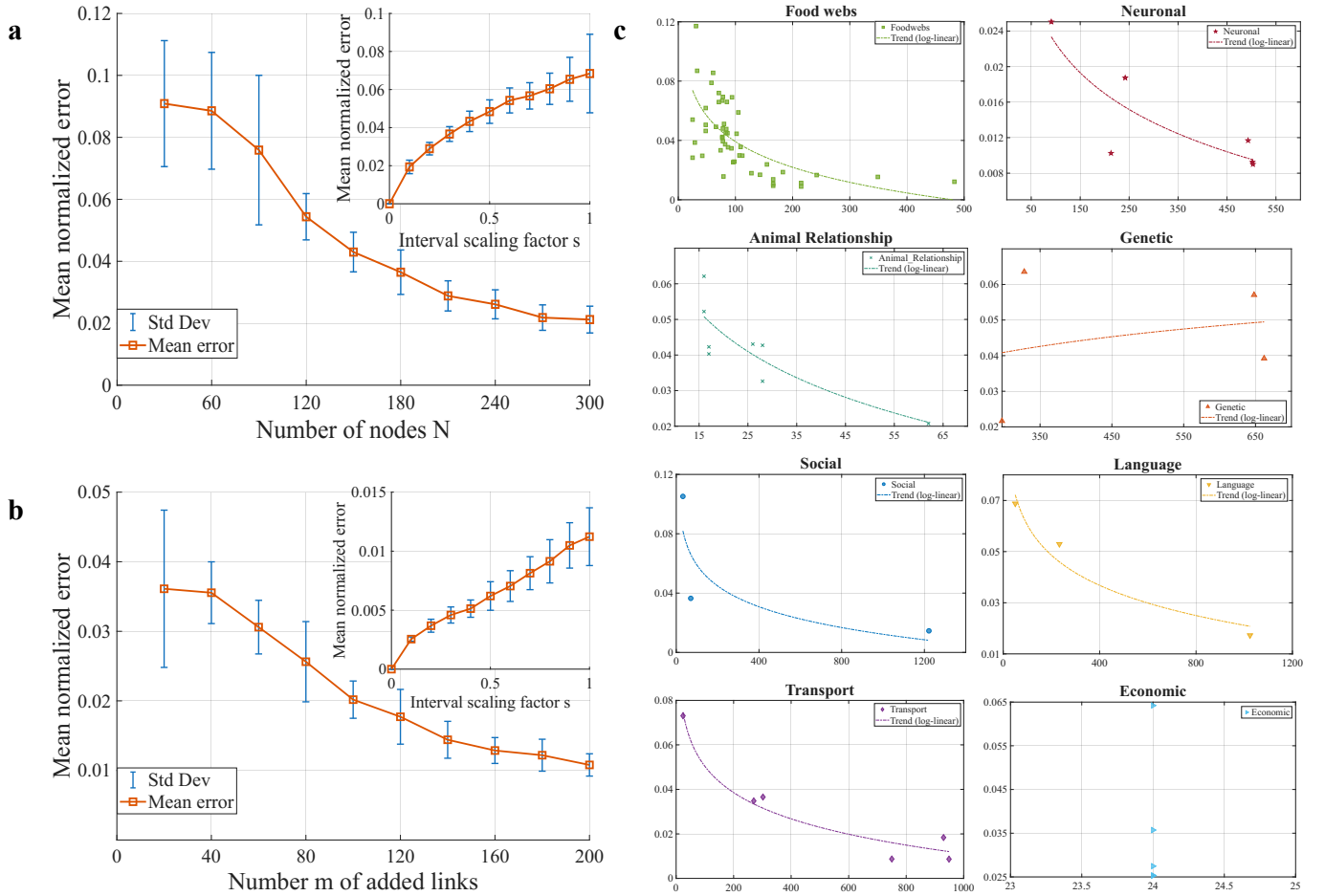}
\caption{\textbf{Accuracy of the spectral reduction and relation to eigenvector localization.}
(\textbf{a}–\textbf{b}) Mean normalized error (MNE) between the full Jacobian $\mathcal{J}_T$ and its localized approximation $\widehat{\mathcal{J}}_\alpha$ under two structural variations: increasing system size $N$ in Erdős–Rényi (ER) networks ($p=0.2$) (\textbf{a}) and increasing the attachment parameter $m$ in Barabási–Albert (BA) networks (\textbf{b}). Each point averages over $50$ independent connected realizations (self-loops excluded) with node-level heterogeneity in both diffusion and reaction parameters, sampled independently from fixed intervals. The error decreases with system size and connectivity, respectively, with error bars indicating one standard deviation. Insets quantify the effect of heterogeneity strength $s$ on fixed ER ($N=90$, $p=0.2$) and BA ($N=500$, $m=m_0=200$) networks, where parameters are sampled from intervals progressively shrunk toward their midpoints ($s=0$ homogeneous, $s=1$ fully heterogeneous). The MNE increases with $s$, indicating reduced accuracy of the approximation as node-to-node variability grows.
(\textbf{c}) MNE for representative real-world networks versus system size, grouped by domain, with log-linear fits indicating an overall decay with size.}
    \label{Fig:Data}
\end{figure*}

We now test empirically the accuracy of the localization reduction by evaluating how well the spectral dynamics can be reconstructed from the host nodes of localized Laplacian eigenvectors. Specifically, we examine how the approximation depends on network size, parameter heterogeneity, and network structure. Erd\H{o}s--R\'enyi (ER) graphs are used as a baseline model because they provide an unconstrained random topology analogous to random matrix ensembles. Figure~\ref{Fig:Data}(a) compares the spectrum of the full system Jacobian $\mathcal{J}_T$ with that predicted by the localized approximation $\widehat{\mathcal{J}}_\alpha$. The discrepancy is quantified through the \emph{mean normalized eigenvalue error} (MNE),
\[
\mathrm{MNE}
=
\frac{1}{|\Lambda|}
\sum_{\lambda\in\Lambda}
\frac{|\lambda-\hat{\lambda}|}{|\lambda|},
\]
where $\Lambda$ denotes the eigenvalues of $\mathcal{J}_T$ and $\hat{\lambda}$ the corresponding eigenvalue of $\widehat{\mathcal{J}}_\alpha$. The MNE decreases as the network size $N$ increases, reflecting the progressive localization of Laplacian eigenvectors in larger networks: as eigenmodes become increasingly concentrated on restricted subsets of nodes, the node-resolved reduction becomes more accurate. Figure~\ref{Fig:Data}(b) shows the corresponding behavior for Barab\'asi--Albert (BA) scale-free networks as the attachment parameter $m$ increases. In this case, the error also decreases systematically, indicating that the reduction remains accurate in increasingly connected scale-free structures.

The insets of Fig.~\ref{Fig:Data}(a--b) quantify the effect of parameter heterogeneity on the localization reduction. Nodewise parameters are sampled from intervals whose width is controlled by $s$. For a parameter $\theta$ with bounds $[\theta_{\min},\theta_{\max}]$, we define its midpoint $\theta_0=(\theta_{\min}+\theta_{\max})/2$ and half-width $\Delta\theta=(\theta_{\max}-\theta_{\min})/2$, and sample nodewise values as
\(
\theta_i=\theta_0+s\,\Delta\theta\,(2\xi_i-1),\quad
\xi_i\sim\mathrm{Unif}(0,1).
\) When $s=0$, all parameters collapse to their midpoint values and the system is homogeneous across nodes. Increasing the scaling factor $s$ increases the node-to-node variability, leading to larger approximation errors.

We finally consider empirical systems. Figure~\ref{Fig:Data}(c) reports the MNE for eight real-world networks spanning different domains, including food webs and genetic networks, neuronal and animal interaction networks, social and language networks, transportation systems, and economic networks. For most network families, the approximation error decreases rapidly with increasing network size, consistent with the behavior observed in synthetic graphs. Remarkably, empirical networks often exhibit even smaller errors than their synthetic counterparts, indicating stronger localization and an enhanced accuracy of the reduction in real-world systems. The genetic networks do not exhibit a clear monotonic trend, and the economic networks correspond to systems of identical size, preventing a size-based comparison. Nevertheless, in both cases the approximation error remains small, indicating that the localized reduction remains accurate across these systems. To quantify the overall tendency, networks with identical size are aggregated and their mean error is fitted by a log-linear model $E\approx \beta_0+\beta_1\log N$, revealing a systematic decrease of the error with system size across the ensemble of networks.

As shown in the Supplementary Material through representative eigenvector heatmaps and additional localization analyses, the improved accuracy observed in BA and empirical networks relative to ER graphs is consistent with stronger spectral localization of the corresponding Laplacian modes. In BA networks, the approximation becomes increasingly accurate as the attachment parameter $m$ increases, while empirical networks display surprisingly strong localization and systematically outperform the ER baseline.

%%%%%%%%%%%%%%%%%%%%%%%%%%%%%%%%%%%%%%%%%%%%%%%%%%%%%%%%%%%%%%%%%%
%%%%%%%%%%%%%%%%%%%%%%%%%%%%%%%%%%%%%%%%%%%%%%%%%%%%%%%%%%%%%%%%%%

\subsection{Emergence of Exotic Patterns}
\label{sec:pattern}

Having established the validity of the localized spectral reduction, we now examine how spectral localization organizes nonlinear collective dynamics beyond instability onset. Near continuous bifurcations, unstable linear modes organize the spatial structure of the emerging nonlinear states~\cite{nakao2014complex,di2018ginzburg,asllani2025pattern}. In the present setting, the homogeneous mode associated with the zero Laplacian eigenvalue generates synchronized limit-cycle oscillations, while the destabilization of localized transversal modes organizes the emergence of heterogeneous collective behavior around this synchronous oscillatory state. For the zero Laplacian eigenvalue, the reduced Jacobian $\widehat{J}_{\alpha}$ coincides with the local Jacobian of each node; we therefore represent these nodewise values with distinct symbols to quantify the real and imaginary parts of the underlying uncoupled node dynamics. As localized transversal modes become unstable, the dynamics progressively disentangles into subsets of nodes governed by distinct localized instability structures. Spectral localization therefore transfers the organization of collective dynamics from abstract spectral coordinates to identifiable subsets of dominant nodes. We investigate how different localized instability structures shape the nonlinear dynamics of the heterogeneous Brusselator system introduced above, giving rise to cluster synchronization, amplitude chimeras, amplitude--phase chimeras, and localized oscillon-like states. Additional examples are provided in the Supplementary Material, including logistic survivability and distributed control in information spreading (Sec.~II), as well as extensions to mixed models of reaction--diffusion systems (Sec.~III).

\begin{figure*}[t!]
\centering
\includegraphics[width=\textwidth]{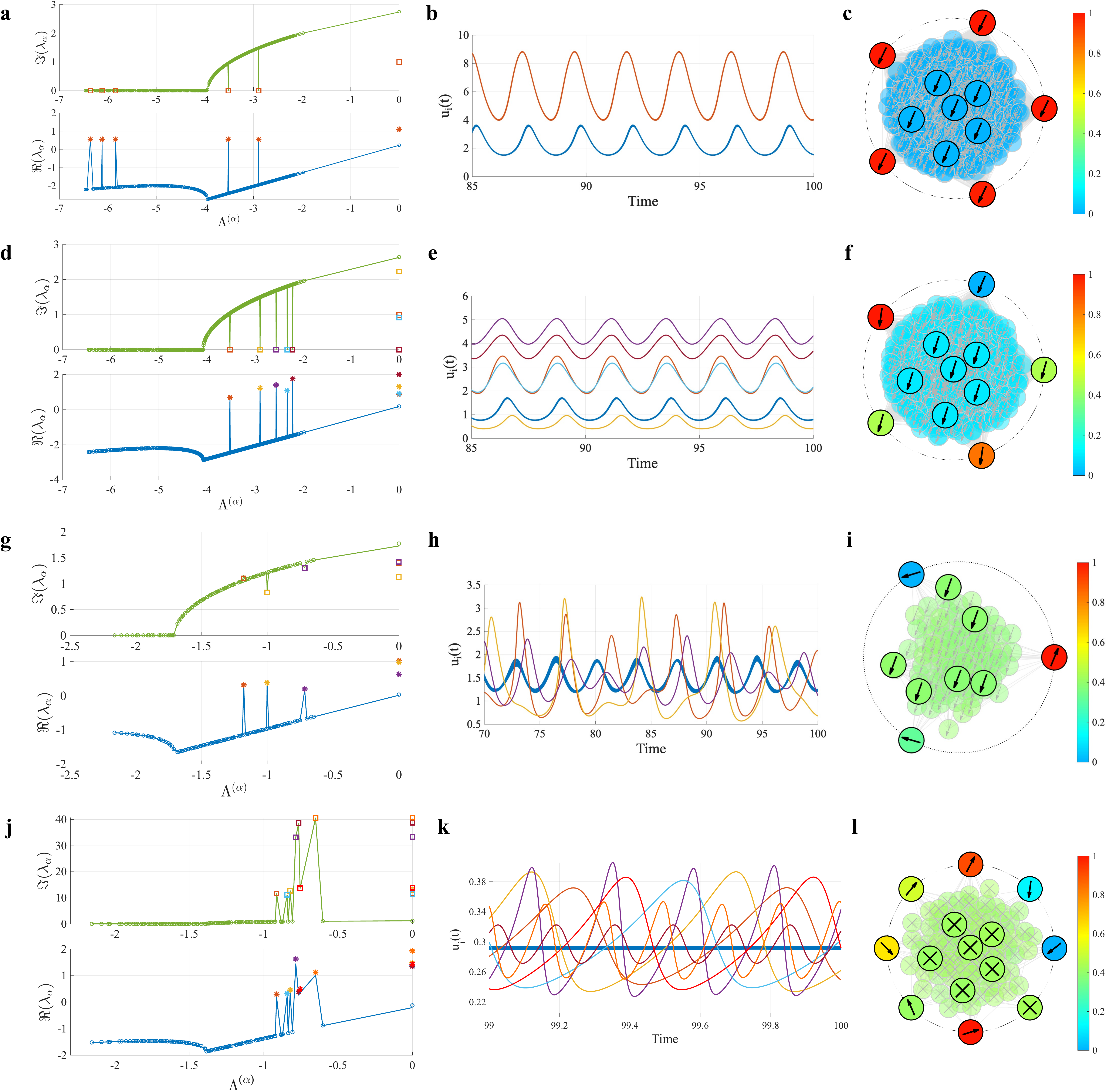}
\caption{\textbf{Spectral localization progressively breaks collective synchronization.}
Each regime is illustrated by the corresponding Master Stability Function (left),
representative node trajectories (middle), and the spatial configuration of the
network (right).
(\textbf{a}--\textbf{c}) \textit{Cluster synchronization}: the transverse spectrum splits into
two groups of exponents, leading to two internally synchronized clusters
that remain frequency-locked through network coupling.
(\textbf{d}--\textbf{f}) \textit{Amplitude chimera}: differences in transverse growth rates generate
persistent amplitude segregation while phases remain aligned.
(\textbf{g}--\textbf{i}) \textit{Amplitude--phase chimera}: unstable modes acquire non-zero imaginary
parts, leading to simultaneous amplitude contrast and phase dispersion.
(\textbf{j}--\textbf{l}) \textit{Oscillon patterns}: strong spectral localization isolates a subset of
oscillatory nodes while the remaining nodes converge to the homogeneous equilibrium.
All parameters and numerical details are provided in the Supplementary Material.
}\label{fig:Fig2}
\end{figure*}

\emph{a) Cluster synchronization.}
In the regime shown in Fig.~\ref{fig:Fig2}(a--c), the transverse spectrum splits into two groups of unstable exponents. Cluster synchronization corresponds to the formation of internally synchronized groups of oscillators that remain dynamically distinct from one another~\cite{arenas_synchronization_2008,belykh2008cluster,sorrentino2016complete}. Nodes associated with the same spectral branch escape with identical growth rates and therefore evolve with the same oscillation amplitude, forming internally synchronized clusters. Within each group, the imaginary parts of the transverse exponents are also identical, ensuring phase synchronization inside the cluster. Although the two groups possess different transverse growth rates, their strong mutual coupling maintains frequency locking across the network. The resulting dynamics organizes into two coherent synchronized clusters, as shown by the trajectories in Fig.~\ref{fig:Fig2}(b) and the spatial configuration in Fig.~\ref{fig:Fig2}(c).

\emph{b) Amplitude chimera.}
A small modification of the transverse spectral structure leads to a qualitatively different collective regime. In Fig.~\ref{fig:Fig2}(d--f), unstable modes retain similar imaginary parts but develop different positive real components. As a consequence, subsets of nodes remain frequency locked while escaping with different growth rates, generating persistent amplitude segregation without phase incoherence. The resulting state is an amplitude chimera, where coherent oscillatory dynamics coexists with stable amplitude differentiation across the network. Unlike classical chimera scenarios arising from imposed asymmetries or parameter mismatches \cite{chimera,Abrams_Strogatz,Panaggio_2015,zakharova_chimera_2020}, the present mechanism emerges directly from the localized transverse spectral hierarchy. 

\emph{c) Amplitude--phase chimera coexistence.}
A further restructuring of the unstable spectral branches produces the regime shown in Fig.~\ref{fig:Fig2}(g--i). In contrast to the amplitude-chimera case, the unstable modes now possess both distinct real parts and non-zero imaginary components. The escaping nodes therefore separate not only in amplitude but also in phase, generating simultaneous amplitude contrast and phase dispersion across the network. Spectral analysis reveals partially overlapping unstable supports and near-degenerate growth rates, allowing multiple localized modes to interact simultaneously. The resulting dynamics combines coherent and incoherent behavior in both amplitude and phase, producing a full amplitude--phase chimera organized by the localized transverse spectral hierarchy.

\emph{d) Oscillon patterns.}
Through the same spectral mechanism underlying the previous regimes, one subset of nodes can be driven into sustained oscillations while the complementary subset remains at the original homogeneous equilibrium. The final regime (Fig.~\ref{fig:Fig2}(j--l)) displays oscillon-like behavior: spatially localized, time-periodic excitations embedded in an otherwise stationary background. Oscillons were first reported in granular and chemical reaction--diffusion systems~\cite{umbanhowar_localized_1996,Vanag_Epstein}, and related localized oscillatory structures have also been identified in neuronal systems and real-world networks~\cite{oscillon_net,muolo_persistence_2024}. In the present setting, however, the mechanism follows directly from the localized transverse spectral hierarchy. A subset of seven nodes projects predominantly onto unstable modes with positive real parts and non-zero imaginary components, leading to localized limit-cycle oscillations, while the remaining nodes align with stable modes and converge toward fixed points. The coexistence of oscillatory and stationary domains places this regime in close relation to chimera death~\cite{amp_chimera}. The oscillatory nodes sustain spatially pinned oscillations (Fig.~\ref{fig:Fig2}(k)), whereas the remaining nodes remain stationary. The unstable localized modes visible in Fig.~\ref{fig:Fig2}(j) are directly reflected in the schematic at the right of Fig.~\ref{fig:Fig2}, where the outer nodes correspond to unstable limit cycles and the inner nodes to stable fixed points.

Together, Fig.~\ref{fig:Fig2} shows that spectral localization provides a unifying mechanism linking instability structure, node-level organization, and nonlinear collective pattern formation in heterogeneous networks.

%%%%%%%%%%%%%%%%%%%%%%%%%%%%%%%%%%%%%%%%%%%%%%%%%%%%%%%%%%%%%%%%%%%%%%%%%%%%%%%%%%%%%%%%%%%%
%%%%%%%%%%%%%%%%%%%%%%%%%%%%%%%%%%%%%%%%%%%%%%%%%%%%%%%%%%%%%%%%%%%%%%%%%%%%%%%%%%%%%%%%%%%%

\section{Conclusions}
\label{sec:end}

Real complex systems are intrinsically heterogeneous. Their components rarely share identical dynamics: nodes may differ in parameters, timescales, interaction strengths, or even in the governing equations themselves. While this diversity is central to biological, social, technological and ecological systems, it also places them beyond the reach of many classical analytical frameworks built on symmetry, identical units, and exact modal decoupling.

Here we showed that the same structural disorder and heterogeneity that obstruct conventional spectral analysis can also restore interpretability. Classical stability theory proceeds by transforming dynamics from physical space into spectral space, where collective transitions are organized by unstable modes. However, these modes generally cannot be directly associated with identifiable physical components. We demonstrated that spectral localization fundamentally alters this picture: as Laplacian eigenvectors localize onto restricted subsets of nodes, the spectral representation approaches an almost node-resolved basis, generating a mode--node correspondence.

Building on this principle, we introduced a localization reduction framework that transforms a high-dimensional heterogeneous stability problem into a set of low-dimensional node-resolved spectral problems. The resulting framework accurately reproduces the heterogeneous Jacobian spectrum, identifies the nodes driving instability, and becomes increasingly effective in large and empirical networks where localization is strongest. 

Beyond instability onset, the localized spectral hierarchy organizes a broad family of nonlinear collective states, including cluster synchronization, amplitude chimeras, amplitude--phase chimeras, and localized oscillon-like patterns. Because localized modes can be independently controlled through their dominant nodes, the framework naturally provides a route toward distributed intervention and control of collective behavior.

More broadly, our results reveal a counterintuitive principle of complex systems: the same heterogeneity and disorder that obstruct exact solvability can also generate the localization structure that restores physical interpretability. In this sense, complexity does not merely obscure collective behavior; under suitable conditions, it can reveal the microscopic components that organize it.

The present framework opens a node-centered perspective on collective dynamics, where local parameters, coupling architecture, and network topology combine to organize the emergence of macroscopic behavior. This perspective may open new directions for the interpretation, prediction, and control of collective phenomena across complex biological, technological, and social systems.

%%%%%%%%%%%%%%%%%%%%%%%%%%%%%%%%%%%%%%%%%%%%%%%%%%%%%%%%%%%%%%%%%%%%%%%%%%%%%%%%%%%%%%%%%%%%
%%%%%%%%%%%%%%%%%%%%%%%%%%%%%%%%%%%%%%%%%%%%%%%%%%%%%%%%%%%%%%%%%%%%%%%%%%%%%%%%%%%%%%%%%%%%

\section{Methods}
\label{sec:appendix}

\subsection{Brusselator Model}
\label{sec:bruss}

The Brusselator was originally introduced as a minimal theoretical model inspired by the Belousov--Zhabotinsky reaction, aimed at capturing the emergence of self-sustained chemical oscillations in nonequilibrium systems \cite{prigogine1968symmetry}. It describes the local kinetics of two interacting species and is governed by
\begin{equation}\label{eq:bruss_aspatial}
\dot{u} = a-(b+d)u + c u^2 v,\qquad
\dot{v} = b u - c u^2 v ,
\end{equation}
remembered for its ability to exhibit a transition from a stable steady state to persistent oscillations as reaction parameters are varied. The system admits a unique equilibrium $u^*={a}/{d}, v^*={b\,d}/{a\,c}$ which is stable for sufficiently small values of the control parameter $b$. Linear stability analysis shows that this equilibrium loses stability through a Hopf bifurcation when the trace of the Jacobian vanishes while the determinant remains positive. This condition yields the critical threshold
\(
b_H = d + c\,{a^2}/{d^2},
\)
beyond which the system develops stable limit-cycle oscillations. These intrinsic oscillations constitute the local dynamical ingredient underlying the network-induced patterns analyzed in the main text.

\subsection{Eigenvalue pairing algorithm}
\label{sec:pairing}

The eigenvalues of the reduced Jacobians $\widehat{\mathcal{J}}_\alpha$ associated with
the transversal Laplacian modes $\Lambda^{(\alpha)}$ are naturally indexed by
the mode label $\alpha\ge 2$, with each mode contributing $M$ eigenvalues.
By contrast, the $MN$ eigenvalues of the full system Jacobian
$\mathcal{J}_T\in\mathbb{C}^{MN\times MN}$ are not canonically organized according to
Laplacian modes. To compare the reduced and full spectra, we construct two sets,
$\mathcal{U}$, obtained by aggregating the eigenvalues of
$\{\widehat{\mathcal{J}}_\alpha\}_{\alpha=2}^N$, and $\mathcal{V}$, containing the
eigenvalues of $\mathcal{J}_T$. Because there is no natural ordering aligning the two
spectra, we determine an injective correspondence by solving a minimum-weight
bipartite matching problem in the complex plane, where the pairing cost is the
complex modulus of the eigenvalue difference. The optimal assignment,
computed using the Hungarian algorithm
\cite{kuhn1955hungarian,munkres1957algorithms},
minimizes the total spectral discrepancy between the reduced and full systems.
With the resulting pairing between $\mathcal{U}$ and a subset of
$\mathcal{V}$, we quantify agreement via the mean normalized error, defined as
the average complex distance between matched eigenvalues, normalized by their
magnitude (with a unit lower bound to ensure robustness near zero). This metric
provides a systematic measure of how accurately the localized transversal
approximation reproduces the full Jacobian spectrum.

%%%%%%%%%%%%%%%%%%%%%%%%%%%%%%%%%%%%%%%%%%%%%%%%%%%%%%%%%%%%%%%%
%%%%%%%%%%%%%%% SUPPLEMENTARY MATERIAL %%%%%%%%%%%%%%%%%%%%%%%%%
%%%%%%%%%%%%%%%%%%%%%%%%%%%%%%%%%%%%%%%%%%%%%%%%%%%%%%%%%%%%%%%%

\clearpage
\onecolumngrid
\appendix

\begin{center}

{\huge \bfseries Supplementary Material}

\vspace{0.3cm}

{\Large \textbf{Complexity Reveals the Microscopic Origins of Macroscopic Dynamics}}

\vspace{0.5cm}

{\large Haoyang Qian$^{1}$,
Beata Casiday$^{2}$,
Gabriel Hood$^{3}$,
Malbor Asllani$^{1}$}

\vspace{0.3cm}

$^{1}$Department of Mathematics, Florida State University, Tallahassee, FL, USA

$^{2}$Yale University, New Haven, CT, USA

$^{3}$Georgia Institute of Technology, Atlanta, GA, USA

%\vspace{0.3cm}

\end{center}

\setcounter{section}{1}
\renewcommand{\thesection}{S}
\renewcommand{\thefigure}{S\arabic{figure}}
\renewcommand{\thetable}{S\arabic{table}}
\setcounter{figure}{0}
\renewcommand{\thefigure}{S\arabic{figure}}

\section*{S1: Localized--Spectral Measures for the graph Laplacian eigenvectors}
\label{sec:measures}

A central ingredient of the localized reduction framework developed in the main text is the spontaneous localization of Laplacian eigenvectors in heterogeneous networks. In many real-world systems, the eigenmodes of the coupling operator are not spatially extended across the entire network, but instead concentrate around a relatively small subset of nodes. This property allows each mode to be associated with a dominant \emph{host node}, providing a simultaneous mode--node decomposition of the dynamics and enabling reduced descriptions based on sparse local information.

\begin{figure}[htbp]
\centering
\includegraphics[width=\textwidth]{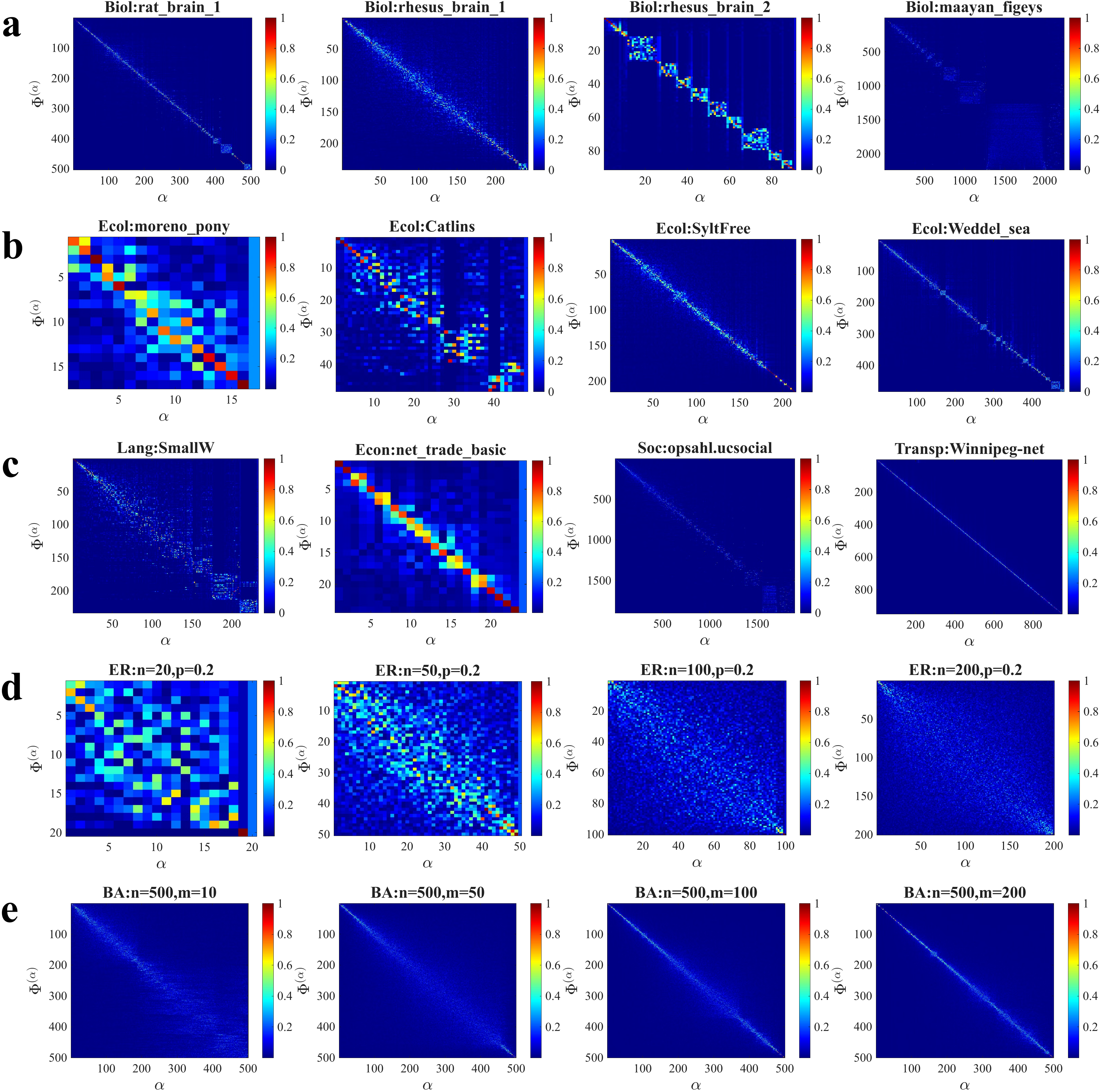}
\caption{
\textbf{Examples of networks from different domains.} 
(\textbf{a}) Biological networks: Mouse’s primary visual cortex connectome 1 \cite{BotaSwanson2007}, 
Connectome of the Rhesus brain, extracted from tract
tracing studies collated in the CoCoMac database \cite{Harriger2012}, 
Connectome of the Rhesus brain, via a retrograde tracer study \cite{Markov2013}, 
and the Human protein-protein interactome produced by a mass spectrometry-based approach by Figeys et al. \cite{Ewing2007}. 
(\textbf{b}) Ecological networks: Dominance among ponies \cite{CluttonBrock1976}, 
River Foodweb in Caitlins Stream, New Zealand \cite{ThompsonTownsend2003}, 
Marine Foodweb in Sylt Tidal Basin, Germany \cite{Dunne2013}, 
and the Marine Foodweb in Weddel Sea, Antarctica \cite{Eklof2013}. 
(\textbf{c}) Language network: Citations from papers that cite ``Small World Problem'' \cite{GarfieldHistCite}. Economic network: International trade network of manufactured goods \cite{DeNooy2018}. Social network: Online messages from an online community of students from the University of California, Irvine \cite{Opsahl2009}. Transportation network: Roads Network in Winnipeg, Canada \cite{TransportationNetworks}. 
(\textbf{d}) Erd\H{o}s--R\'enyi random graphs with connection probability $p=0.2$ and different sizes ($n=20,50,100,200$). 
(\textbf{e}) Networks generated using the Barabási--Albert model, starting from an initial random graph $G(m_0=200, p=0.1)$, for different attachment parameters $m = 10, 50, 100, 200$.
}
\label{fig:twenty-sub}
\end{figure}

To systematically characterize localization across empirical and synthetic networks, we compute several complementary mode-level localization measures for the Laplacian eigenvectors. Figure~\ref{fig:twenty-sub} shows representative examples of the network families considered throughout this work, including biological, ecological, social, transportation, and random graph models. These networks exhibit markedly different spectral organizations, ranging from highly localized structures to more spatially extended modes.

Let $\boldsymbol{\phi}^{(\alpha)}$ be a normalized eigenvector of the network Laplacian $\mathcal{L}$, satisfying
\(
\sum_{i=1}^N \left|\phi_i^{(\alpha)}\right|^2 = 1 .
\)
We define the node weight of mode $\alpha$ by
\begin{equation}
w_i^{(\alpha)} = \left|\phi_i^{(\alpha)}\right|^2 ,
\qquad 
\sum_{i=1}^N w_i^{(\alpha)} = 1 .
\end{equation}
The inverse participation ratio is then given by
\begin{equation}
    \mathrm{IPR}^{(\alpha)}
=
\sum_{i=1}^N \left(w_i^{(\alpha)}\right)^2
=
\sum_{i=1}^N \left|\phi_i^{(\alpha)}\right|^4 .
\end{equation}
A larger value of $\mathrm{IPR}^{(\alpha)}$ indicates that the eigenvector mass is concentrated on fewer nodes, while a smaller value corresponds to a more delocalized mode. For comparison across networks of different sizes, we also use the normalized inverse participation ratio
\begin{equation}
\mathrm{IPR}_{\mathrm{norm}}^{(\alpha)}
=
\frac{\mathrm{IPR}^{(\alpha)} - 1/N}{1 - 1/N},
\end{equation}
which maps the fully delocalized case to $0$ and the fully localized case to $1$.

The effective number of participating nodes is defined as
\begin{equation}
N_{\mathrm{eff}}^{(\alpha)}
=
\frac{1}{\mathrm{IPR}_{\mathrm{norm}}^{(\alpha)}} .
\end{equation}
This quantity estimates the number of nodes that effectively support the eigenvector. Therefore, a smaller $N_{\mathrm{eff}}^{(\alpha)}$ indicates stronger localization. However, networks with $N_{\mathrm{eff}}^{(\alpha)}>100$ tend to produce relatively large approximation errors, which weakens the monotone trend between $N_{\mathrm{eff}}^{(\alpha)}$ and the normalized error.

While the IPR and effective participation number quantify localization globally, the localized reduction itself relies on the existence of dominant host nodes carrying a large fraction of the eigenvector mass. To characterize this feature more directly, we additionally compute peak and tail localization measures.

We define the peak weight of the eigenvector by
\begin{equation}
p^{(\alpha)}
=
\max_{1 \leq i \leq N} w_i^{(\alpha)}
=
\max_{1 \leq i \leq N} \left|\phi_i^{(\alpha)}\right|^2 .
\end{equation}
This quantity measures the weight carried by the dominant node of the eigenvector. In our localized mode approximation, the host node of mode $\alpha$ is chosen as
\[
\eta(\alpha)
=
\arg\max_{1 \leq i \leq N} \left|\phi_i^{(\alpha)}\right| .
\]
Thus, $p^{(\alpha)}$ directly quantifies how much of the eigenvector is concentrated on the selected host node. Finally, we define the tail weight by
\begin{equation}
    \tau^{(\alpha)}
=
1 - p^{(\alpha)} .
\end{equation}
The tail weight measures the total eigenvector mass outside the dominant node. A smaller value of $\tau^{(\alpha)}$ indicates that the mode is more accurately represented by its host node alone. In the numerical results, these quantities are computed for all nontrivial Laplacian eigenvectors, excluding the uniform zero mode, and then averaged over modes to summarize the overall localization strength of each network.

\begin{figure}[t]
\centering
\includegraphics[width=\linewidth]{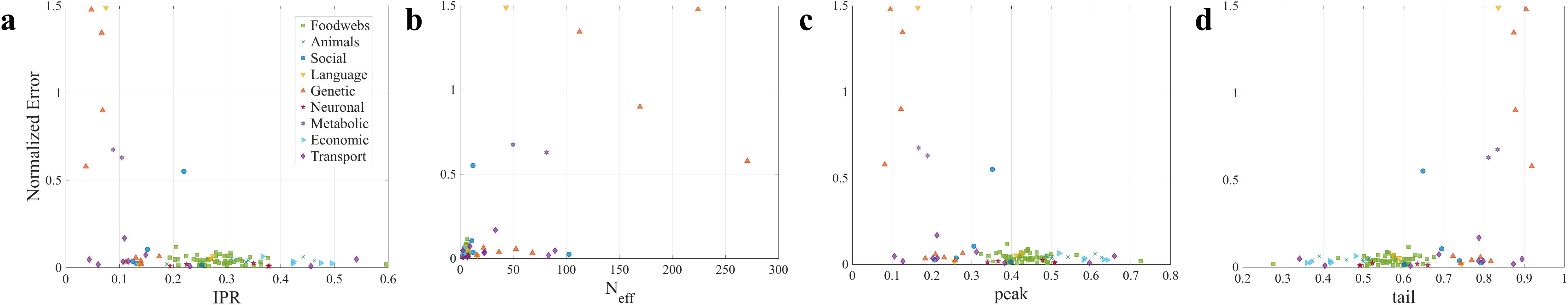}
\caption{\textbf{Normalized eigenvalue approximation error as a function of different eigenvector localization measures for real-world networks from different domains.} Each point represents one network, with colors and markers indicating the corresponding network category. Panels show the normalized error versus (\textbf{a}) the normalized inverse participation ratio (IPR), (\textbf{b}) the effective number of participating nodes $N_{\mathrm{eff}}$, (\textbf{c}) the peak weight on the dominant node, and (\textbf{d}) the tail weight outside the dominant node.}
\label{fig:localization_error}
\end{figure}

Figure~\ref{fig:localization_error} compares the normalized eigenvalue approximation error with four localization measures across real-world networks from different domains. In panel (a), larger IPR values generally correspond to smaller errors, suggesting that stronger eigenvector localization improves the accuracy of the localized approximation. A similar trend is observed in panel (c), where networks with larger peak weights tend to have smaller errors, indicating that modes concentrated on a dominant host node are better captured by the host-node reduction. Panel (d) shows the same effect from the opposite perspective: smaller tail weights, meaning less eigenvector mass outside the dominant node, are associated with smaller approximation errors. Panel (b) shows the relationship with $N_{\mathrm{eff}}$, where smaller values indicate stronger localization. Although smaller $N_{\mathrm{eff}}$ is generally expected to reduce the error, several networks with $N_{\mathrm{eff}}>100$ produce relatively large errors, so this measure does not display as clear a monotone trend as the other localization measures.

\section*{S2: Further applications}

\subsection{Logistic survivability}

As a concrete illustration of the localization reduction method for analysing dynamics in heterogeneous empirical systems, we consider the trade network shown in Fig.~\ref{fig:Fig1}. Despite its modest size, this network already exhibits pronounced Laplacian eigenvector localization, making it a suitable testbed for the framework. To keep the dynamical setting fully transparent, we endow the network with a minimal heterogeneous logistic dynamics~\cite{murray_mathematical_2002}
\begin{equation}\label{eq:hetero_logistic}
\dot{x}_i
=
r_i x_i(1-x_i)
+
D_i\sum_{j=1}^N L_{ij}x_j,
\quad i=1,\dots,N,
\end{equation}
where \(x_i\) is a generic node activity (such as production or traded volume), while \(r_i\) and \(D_i\) are node-dependent growth and coupling parameters. The logistic term captures self-limited local expansion under finite capacity or market saturation, and the Laplacian term describes exchange through network links. The inactive state \(x_i^*=0\) is always an equilibrium. The model is not intended as a literal description of trade dynamics, but as a one-variable prototype that allows the localization reduction method to be demonstrated analytically and interpreted directly at the node level.
Linearizing around this reference state yields
\begin{equation}\label{eq:linearized_hetero_logistic}
\dot{\boldsymbol{\xi}} = \mathcal{J}\boldsymbol{\xi},
\qquad
\mathcal{J}=\operatorname{diag}(r)+\operatorname{diag}(D)\mathcal{L},
\end{equation}
where \(\boldsymbol{\xi}=(\xi_1,\dots,\xi_N)^\top\) is the perturbation vector, \(\operatorname{diag}(r)=\operatorname{diag}(r_1,\dots,r_N)\), and \(\operatorname{diag}(D)=\operatorname{diag}(D_1,\dots,D_N)\). This is the scalar (\(M=1\)) counterpart of the general linearized operator of the main text, simplified by the one-variable node dynamics and by the fact that coupling is already written in Laplacian form.
The linearized operator is therefore diagonalized by the Laplacian eigenbasis. Each localized mode \(\alpha\) inherits the parameters of its host node \(\eta(\alpha)\), and the localized Master Stability Function reduces to
\begin{equation}\label{eq:MSF_hom}
\Psi\left(\Lambda^{(\alpha)},\eta(\alpha)\right)
:=
\lambda_\mathcal{J}^{(\alpha)}
=
r_{\eta(\alpha)}+D_{\eta(\alpha)}\Lambda^{(\alpha)}.
\end{equation}
This approximation provides a direct link between spectral localization and nodewise instability thresholds: instability is controlled not only by parameter values, but also by where the corresponding modes are concentrated in the network.

\begin{figure*}[t]
\centering
\includegraphics[width=\textwidth]{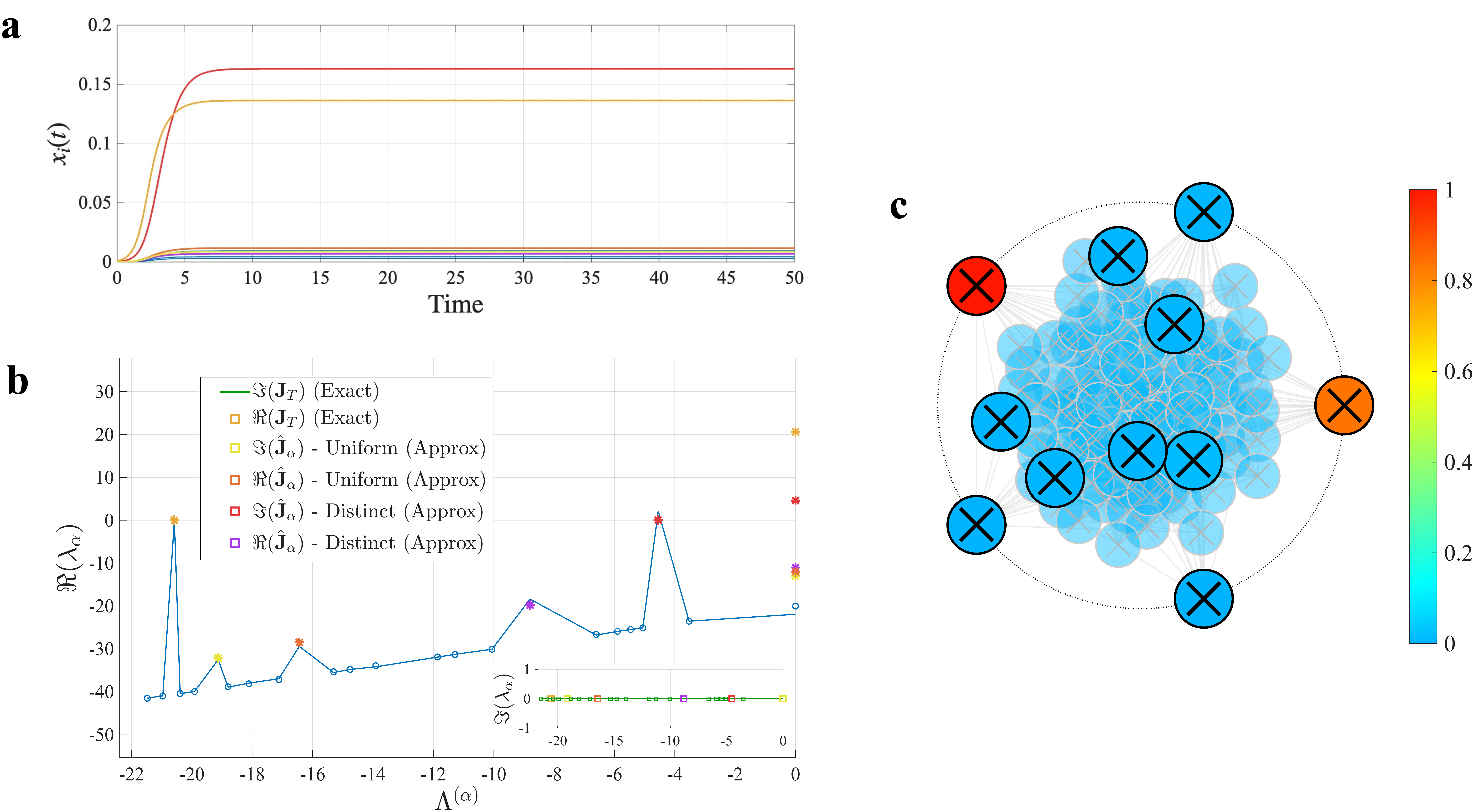}
\caption{
\textbf{Individual dynamics disentanglement in the logistic model.}
(\textbf{a}) Time series of $x_i(t)$ for a network realization in which all nodes share the same equilibrium point; two trajectories separate from the bulk, consistent with an instability associated with positive real-part eigenvalues.
(\textbf{b}) Master-stability-style spectrum comparison for the logistic model $\dot{x}_i = r_i x_i(1-x_i) + D \sum_j L_{ij}x_j$
on the trade network. Solid lines denote eigenvalues of the full-system Jacobian $\mathcal{J}_T$, while circles denote the eigenvalues of the localized approximation $\widehat{\mathcal{J}}_\alpha$. The five color-coded modes correspond to five nodes with heterogeneous parameters but identical equilibrium points. Among these, two eigenvalues satisfy $\Re(\lambda)>0$, indicating linear instability. Here $D=1$ for all nodes; for the five highlighted nodes,
$r=(20.6,-13,-12,-11,4.6)$, while all remaining nodes have $r=-20$.
(\textbf{c}) Final-state visualization of $x_i(t)$, where the five outer-ring nodes correspond to the highlighted nodes in panel. }
\label{fig:Fig1}
\end{figure*}

This mechanism is confirmed in Fig.~\ref{fig:Fig1}(a,b). The localized MSF in panel (b) predicts that only two modes-nodes acquire positive growth rates, each tied to the nodes on which the corresponding eigenvectors are concentrated. In agreement, the time series in panel (a) show that activity re-emerges almost exclusively on those nodes, while the remainder stay near the inactive state. This contrasts with the collective behavior expected in the homogeneous counterpart of the system, where identical parameters make the uniform mode lose stability first and drive coherent growth across all nodes. Here, by contrast, competition among heterogeneous local growth rates breaks that global response. Instability no longer unfolds through a purely modal hierarchy, but through a simultaneous mode–node decomposition in which different parts of the network follow distinct routes away from the inactive state. In the trade-network setting, this corresponds to renewed production or expansion of manufactured-goods exchanges arising in selected markets, while much of the system remains dormant.

\subsection{Distributed controlled in information spreading}
\label{sec:SIS}

\begin{figure*}[t!]
    \includegraphics[width=\textwidth]{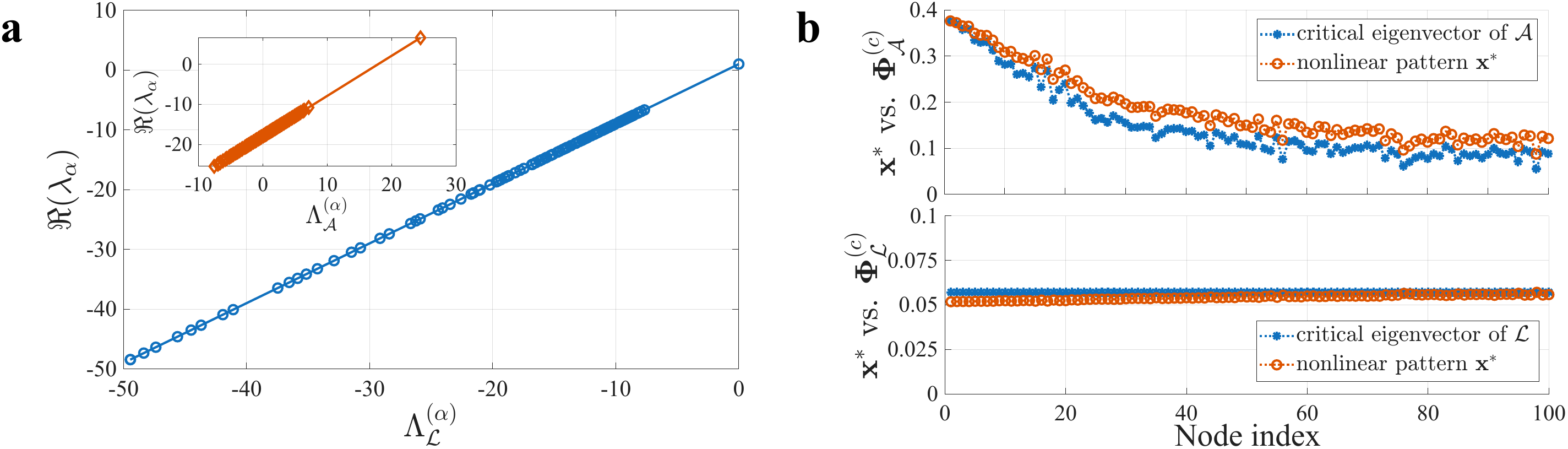}
\caption{\textbf{Comparison between adjacency-based and Laplacian-based contagion dynamics.} (\textbf{a}) Master stability spectra for the two formulations. The inset shows the adjacency-based case, where growth rates are controlled by the eigenvalues of \(\mathcal{A}\). The main panel shows the Laplacian-based case, where the spectrum is organized by the eigenvalues of \(\mathcal{L}\). (\textbf{b}) Steady-state activity profile \(x^*\) compared with the dominant mode in each formulation. Top: \(x_i^*\) against the critical eigenvector of \(\mathcal{A}\), showing a heterogeneous pattern concentrated on structurally favoured nodes. Bottom: \(x_i^*\) against the critical eigenvector of \(\mathcal{L}\), showing an approximately uniform activation profile aligned with the collective Laplacian mode.}
\label{fig:Fig3}
\end{figure*}

As a second illustration, we consider a contagion process on a network. In its standard adjacency-based form, the SIS dynamics reads
\begin{equation}
\dot{x}_i=-\gamma x_i+\beta (1-x_i)\sum_{j=1}^N \mathcal{A}_{ij}x_j,
\qquad i=1,\dots,N,
\end{equation}
where \(x_i(t)\) denotes the activity level at node \(i\), interpreted for instance as infection prevalence, adoption probability, or opinion activation. The parameters \(\gamma\) and \(\beta\) are the recovery and transmission rates, respectively. Linearization about the inactive state \(x_i^*=0\) gives
\begin{equation*}
\dot{\bm{\xi}} = (-\gamma \mathbb{I}+\beta \mathcal{A})\,\bm{\xi}.
\end{equation*}
If \(\mathcal{A}\bm{\Phi}_\mathcal{A}^{(\alpha)}=\Lambda_\mathcal{A}^{(\alpha)}\bm{\Phi}_\mathcal{A}^{(\alpha)}\), the corresponding master stability function is
\begin{equation*}
\Psi\left(\Lambda_\mathcal{A}^{(\alpha)}\right)=-\gamma+\beta \Lambda_\mathcal{A}^{(\alpha)}.
\end{equation*}
The resulting spectrum is shown in the inset of Fig. \ref{fig:Fig3}(a). Instability is controlled by the leading adjacency eigenmode, so the emerging activity follows a structurally heterogeneous direction. This is confirmed in Fig. \ref{fig:Fig3}(b, top), where the steady state closely aligns with the critical eigenvector of $\mathcal{A}$.

In many spreading settings, however, one may instead wish to favour a more homogeneous activation profile. This is naturally relevant not only for epidemics under compensatory interventions, but also for information or opinion dynamics where broad coverage is desirable. To expose such a mechanism, we rewrite the interaction term in Laplacian form $\mathcal{L}$, so that the dynamics becomes
\begin{equation}
\dot{x}_i=-\gamma_i x_i+\beta(1-x_i)\left(\sum_{j=1}^N \mathcal{L}_{ij}x_j+k_i x_i\right),
\quad i=1,\dots,N,
\end{equation}
where the recovery rate is now allowed to vary across nodes. Linearizing around the inactive state gives
\[
\dot{\boldsymbol{\xi}}
=
\left(-\mathrm{diag}(\gamma_i)+\beta\,\mathrm{diag}(k_i)+\beta \mathcal{L}\right)\boldsymbol{\xi}.
\]
Choosing the node-dependent rates such that \(-\gamma_i+\beta k_i=c_i\), the degree dependence is absorbed into a residual diagonal heterogeneity, yielding
\[
\dot{\boldsymbol{\xi}}=\left(\mathrm{diag}(c_i)+\beta\mathcal{L}\right)\boldsymbol{\xi}.
\]
The system is no longer exactly diagonal in the Laplacian basis, but the localization reduction applies directly: each localized mode inherits the coefficient of its host node. Thus, if \(\mathcal{L}\bm{\Phi}_\mathcal{L}^{(\alpha)}=\Lambda_\mathcal{L}^{(\alpha)}\bm{\Phi}_\mathcal{L}^{(\alpha)}\), the localized Master Stability Function is
\begin{equation}
\Psi\!\left(\Lambda_\mathcal{L}^{(\alpha)},\eta(\alpha)\right)
=
c_{\eta(\alpha)}+\beta\Lambda_\mathcal{L}^{(\alpha)}.
\label{eq:MSF_SIS}
\end{equation}

For the illustrative realization shown in Fig.~\ref{fig:Fig3}, we choose the special case \(c_i\equiv c\), so that the localized expression coincides with the exact Laplacian spectrum. The corresponding spectrum is reported in Fig.~\ref{fig:Fig3}(a, main panel). The dominant mode is now the Laplacian eigenvector associated with \(\Lambda_\mathcal{L}^{(c)}=0\), namely the uniform direction \(\bm{1}=(1,\dots,1)^T\) for a connected network. The onset of spreading therefore aligns with a collective mode that is spatially homogeneous, rather than with a structurally localized adjacency eigenvector. Accordingly, Fig. \ref{fig:Fig3}(b, bottom) shows an approximately uniform steady-state profile. In this way, degree-dependent recovery rates provide a simple mechanism through which heterogeneous networks can be steered from localized outbreaks toward approximately uniform contagion or adoption patterns.

%%%%%%%%%%%%%%%%%%%%%%%%%%%%%%%%%%%%%%%%%%%%%%%%%%%%%%%%%%%%%%%%%%%%%%%%%%%%%%%%%%%%%%%%%%%%
%%%%%%%%%%%%%%%%%%%%%%%%%%%%%%%%%%%%%%%%%%%%%%%%%%%%%%%%%%%%%%%%%%%%%%%%%%%%%%%%%%%%%%%%%%%%

\section*{S3: Beyond Model Constraint}

The localized reduction framework developed in the main text does not rely on a specific reaction--diffusion model, but rather on the spectral organization of the coupling operator and the localization properties of the corresponding Laplacian eigenvectors. To illustrate this robustness beyond identical node dynamics, we consider a heterogeneous network composed of two distinct reaction--diffusion systems coupled through the same Laplacian structure. In particular, we combine Brusselator and Mimura--Murray nodes within a single network and investigate whether the same classes of collective patterns persist under mixed local kinetics.

The Mimura--Murray model was introduced as a minimal predator--prey reaction
system capable of exhibiting oscillatory dynamics through nonlinear species
interactions \cite{mimura1978diffusive}. In its aspatial form, the local kinetics of the prey density
$\phi$ and predator density $\psi$ is governed by
\begin{equation}
\dot{\phi}=\phi\left(\frac{a+b\phi-\phi^2}{c}-\psi\right),\qquad
\dot{\psi}=\psi\big(\phi-(1+d\psi)\big),
\label{eq:mm_aspatial}
\end{equation}
with all parameters positive. Besides trivial equilibria, the system admits a
positive steady state
\begin{equation}
\phi^*
=\frac{(bd-c)+\sqrt{(bd-c)^2+4d(c+ad)}}{2d},
\qquad
\psi^*=\frac{\phi^*-1}{d}.
\label{eq:mm_fp}
\end{equation}
Linear stability analysis shows that this equilibrium loses stability through
a Hopf bifurcation when the trace of the Jacobian vanishes while the determinant
remains positive. This condition yields the critical threshold
\(
c_H={\phi^*(b-2\phi^*)}/{d\psi^*}.
\)
Beyond this threshold the system develops stable limit-cycle oscillations.

\begin{figure*}[t]
\centering
\includegraphics[width=\textwidth]{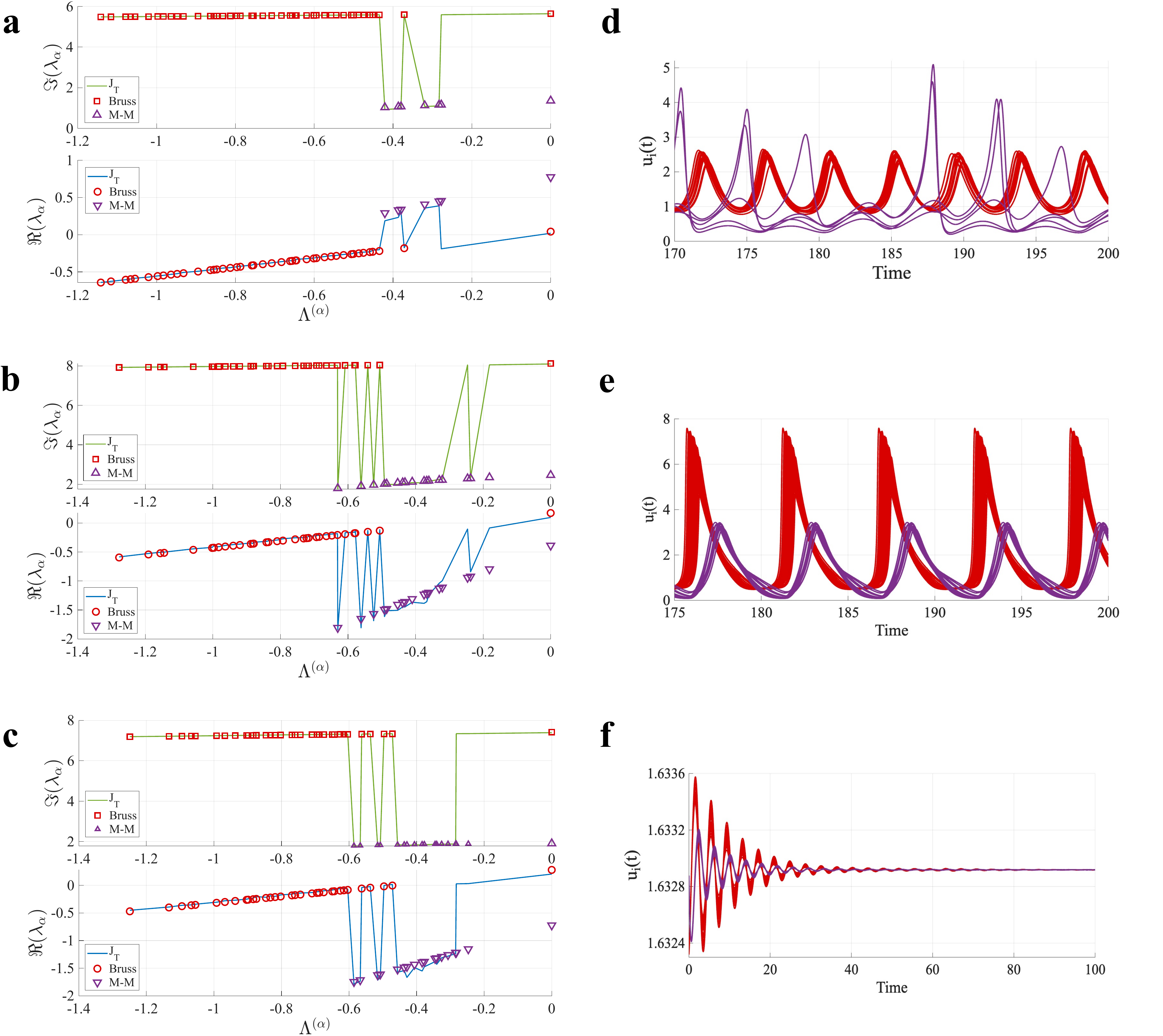}
\caption{\textbf{Dynamics of mixed reaction--diffusion models on the same network.} 
Red quantities correspond to the Brusselator model, while purple quantities correspond to the Mimura--Murray model. Panels (\textbf{a})--(\textbf{c}) show the Master Stability Function together with the eigenvalues of the exact Jacobian $\mathcal{J}_T$ and the localized reduced operator $\widehat{\mathcal{J}}_\alpha$, illustrating the agreement between the full spectral dynamics and the localization-based approximation. Panels (\textbf{d})--(\textbf{f}) display the corresponding node dynamics $u_i(t)$ associated with the unstable modes. 
(\textbf{a},\textbf{d}) Chaotic chimera, characterized by coexistence of coherent and irregular oscillatory regions; 
(\textbf{b},\textbf{e}) cluster synchronization, where groups of nodes evolve coherently; 
and (\textbf{c},\textbf{f}) oscillation death, corresponding to stationary heterogeneous states emerging from the localized instability structure.}
\label{fig:Fig5}
\end{figure*}

We partition the network into two subpopulations governed by distinct local kinetics but coupled through the same Laplacian $L$. 
The first group follows the Brusselator model with state $(u_i^{\mathrm B},v_i^{\mathrm B})$, parameters $(a^{\mathrm B},b^{\mathrm B},c^{\mathrm B},d^{\mathrm B})$, and diffusion $(D^{\mathrm B}_u,D^{\mathrm B}_v)$ (constant across the Brusselator group). 
The second group follows the \emph{Mimura--Murray (MM)} model with state $(u_i^{\mathrm M},v_i^{\mathrm M})$, parameters $(a^{\mathrm M},b^{\mathrm M},c^{\mathrm M},d^{\mathrm M})$, and diffusion $(D^{\mathrm M}_u,D^{\mathrm M}_v)$ (constant across the MM group).

The Brusselator model is given by:
\begin{eqnarray}\label{eq:bruss_two}
\dot{u}_i^{\mathrm B}=a^{\mathrm B}-(b^{\mathrm B}+d^{\mathrm B})u_i^{\mathrm B}
+ c^{\mathrm B}(u_i^{\mathrm B})^2 v_i^{\mathrm B}
+ D^{\mathrm B}_u \sum_{j=1}^N L_{ij}u_j^{\mathrm B},
\\
\dot{v}_i^{\mathrm B}=b^{\mathrm B}u_i^{\mathrm B}
- c^{\mathrm B}(u_i^{\mathrm B})^2 v_i^{\mathrm B}
+ D^{\mathrm B}_v \sum_{j=1}^N L_{ij}v_j^{\mathrm B}.
\end{eqnarray}
The aspatial equilibrium is\,\,
\(
u_{\mathrm B}^*={a^{\mathrm B}}/{d^{\mathrm B}},\quad
v_{\mathrm B}^*={b^{\mathrm B} d^{\mathrm B}}/{a^{\mathrm B} c^{\mathrm B}}.
\)

\begin{table}[t]
\centering
\small
\caption{Parameter sets grouped by node classes (all values rounded to 4 decimals). Distinct parameter tuples define distinct groups.}
\label{tab:group_params}
\begin{tabular}{@{}lrrrrrr@{}}
\toprule
Case / Group & $a$ & $b$ & $c$ & $d$ & $D_u$ & $D_v$ \\
\midrule

\multicolumn{7}{@{}l}{\textbf{Two Model: Cluster}}\\
Brusselator (35 nodes)   & 11.3964 & 18.0104 & 2.7221 & 5.3598 & 0.5000 & 0.7000 \\
Mimura–Murray (15 nodes)     & 3.9746 & 4.7571 & 3.0751 & 0.3620 & 0.5000 & 4.0000 \\
\addlinespace

\multicolumn{7}{@{}l}{\textbf{Two Model: Oscillation Death}}\\
Brusselator (35 nodes)   & 6.8825 & 17.8565 & 4.9064 & 4.2149 & 0.5000 & 0.7000 \\
Mimura–Murray (15 nodes)     & 4.8332 & 2.0397 & 2.4666 & 0.2840 & 0.5000 & 3.0000 \\
\addlinespace

\multicolumn{7}{@{}l}{\textbf{Two Model: Chaotic}}\\
Brusselator (45 nodes)   & 6.1659 & 11.9815 & 3.4161 & 4.0742 & 0.5000 & 0.7000 \\
Mimura–Murray (5 nodes)     & 0.7816 & 19.5058 & 12.0864 & 0.2215 & 0.5000 & 1.8000 \\

\bottomrule
\end{tabular}
\end{table}
\normalsize 

Mimura--Murray model:
\begin{eqnarray}
\label{eq:mm_two}
\dot u_i^{\mathrm M}=u_i^{\mathrm M}\!\left(\frac{a^{\mathrm M}+b^{\mathrm M}u_i^{\mathrm M}-(u_i^{\mathrm M})^2}{c^{\mathrm M}}-v_i^{\mathrm M}\right)
+ D^{\mathrm M}_u \sum_{j=1}^N L_{ij}u_j^{\mathrm M},
\\
\dot v_i^{\mathrm M}=v_i^{\mathrm M}\!\Big(u_i^{\mathrm M}-(1+d^{\mathrm M}v_i^{\mathrm M})\Big)
+ D^{\mathrm M}_v \sum_{j=1}^N L_{ij}v_j^{\mathrm M}.
\end{eqnarray}
Its positive aspatial equilibrium satisfies
\begin{equation}\label{eq:uvstar_M}
u_{\mathrm M}^*
=\frac{(b^{\mathrm M} d^{\mathrm M}-c^{\mathrm M})
+\sqrt{(b^{\mathrm M} d^{\mathrm M}-c^{\mathrm M})^2+4 d^{\mathrm M}(c^{\mathrm M}+a^{\mathrm M} d^{\mathrm M})}}
{2 d^{\mathrm M}},
\qquad
v_{\mathrm M}^*=\frac{u_{\mathrm M}^*-1}{d^{\mathrm M}}.
\end{equation}

We first fix a single MM parameter set $(a^{\mathrm M},b^{\mathrm M},c^{\mathrm M},d^{\mathrm M})$ and compute its equilibrium $(u_{\mathrm M}^*,v_{\mathrm M}^*)$ from the formulas above. 
We then choose the Brusselator parameters so that the Brusselator equilibrium equals the same pair:
pick any $c^{\mathrm B}>0$ and $d^{\mathrm B}>0$, and set\,\,
\(
a^{\mathrm B}=u_{\mathrm M}^*\,d^{\mathrm B},\quad 
b^{\mathrm B}=u_{\mathrm M}^*\,c^{\mathrm B}\,v_{\mathrm M}^*,
\)
which enforces, identically,\,\,
\(
u_{\mathrm B}^*={a^{\mathrm B}}/{d^{\mathrm B}}=u_{\mathrm M}^*,
\quad
v_{\mathrm B}^*={b^{\mathrm B} d^{\mathrm B}}/{a^{\mathrm B} c^{\mathrm B}}=v_{\mathrm M}^* .
\)
(Here diffusion pairs $(D^{\mathrm B}_u,D^{\mathrm B}_v)$ and $(D^{\mathrm M}_u,D^{\mathrm M}_v)$ are model-specific and remain independent; typically $D_u\neq D_v$.)

Figure~\ref{fig:Fig5} shows the resulting dynamics for the mixed-model network. Despite the coexistence of two fundamentally different local reaction mechanisms, the collective behavior remains strongly organized by the Laplacian spectral structure. Panels ({a})--({c}) compare the eigenvalues of the full Jacobian $\mathcal{J}_T$ with the reduced localized approximation $\widehat{\mathcal{J}}_\alpha$, while panels ({d})--({f}) display the corresponding node dynamics. The system exhibits chaotic chimera states, cluster synchronization, and oscillation death, demonstrating that the localization-based reduction remains effective even when the node dynamics are heterogeneous and model-dependent.

These results show that the emergence of collective patterns is governed primarily by the Laplacian modes and their localization properties rather than by the precise form of the local kinetics. In particular, near instability onset, the nonlinear dynamics inherit the spatial organization of the critical localized eigenvectors, allowing the localized reduction framework to remain predictive even beyond identical-model constraints.

%%%%%%%%%%%%%%%%%%%%%%%%%%%%
%%%%%%%%%%%%%%%%%%%%%%%%%%%%

\section*{S4: Identical Hopf conditions, different limit cycles}

To demonstrate that the same equilibrium point does not uniquely determine the nonlinear oscillatory dynamics, we consider two uncoupled single-node Brusselator systems with different parameter sets chosen to share the same steady state 
\(
u^*={a}/{d}, \quad v^*={bd}/{ac}.
\)
We first generate one admissible parameter set \((a_1,b_1,c_1,d_1)\), and then construct a second set \((a_2,b_2,c_2,d_2)\neq (a_1,b_1,c_1,d_1)\) under the constraints
\[
\frac{a_1}{d_1}=\frac{a_2}{d_2}, \qquad \frac{b_1d_1}{a_1c_1}=\frac{b_2d_2}{a_2c_2},
\]
so that both systems have the same equilibrium point. In the example shown here, the two parameter sets are
\(
(a,b,c,d)=(1.2955,\,8.2066,\,4.9526,\,4.3203)
\)
and
\(
(a,b,c,d)=(1.9522,\,9.9884,\,6.0280,\,6.5102),
\)
which yield the common equilibrium
\(
(u^*,v^*)\approx(0.2999,\,5.5260).
\)
We then simulate the corresponding single-node dynamics and compare the resulting trajectories \(u(t)\).

However, the simulated time series \(u(t)\) reveal different limit cycles for the two nodes. This confirms that even when the fixed point is preserved, differences in the underlying kinetic parameters remain sufficient to alter the nonlinear periodic dynamics, so the equilibrium location alone is not enough to characterize the full behavior of the system.

\begin{figure}[h]
    \centering
    \includegraphics[width=.5\columnwidth]{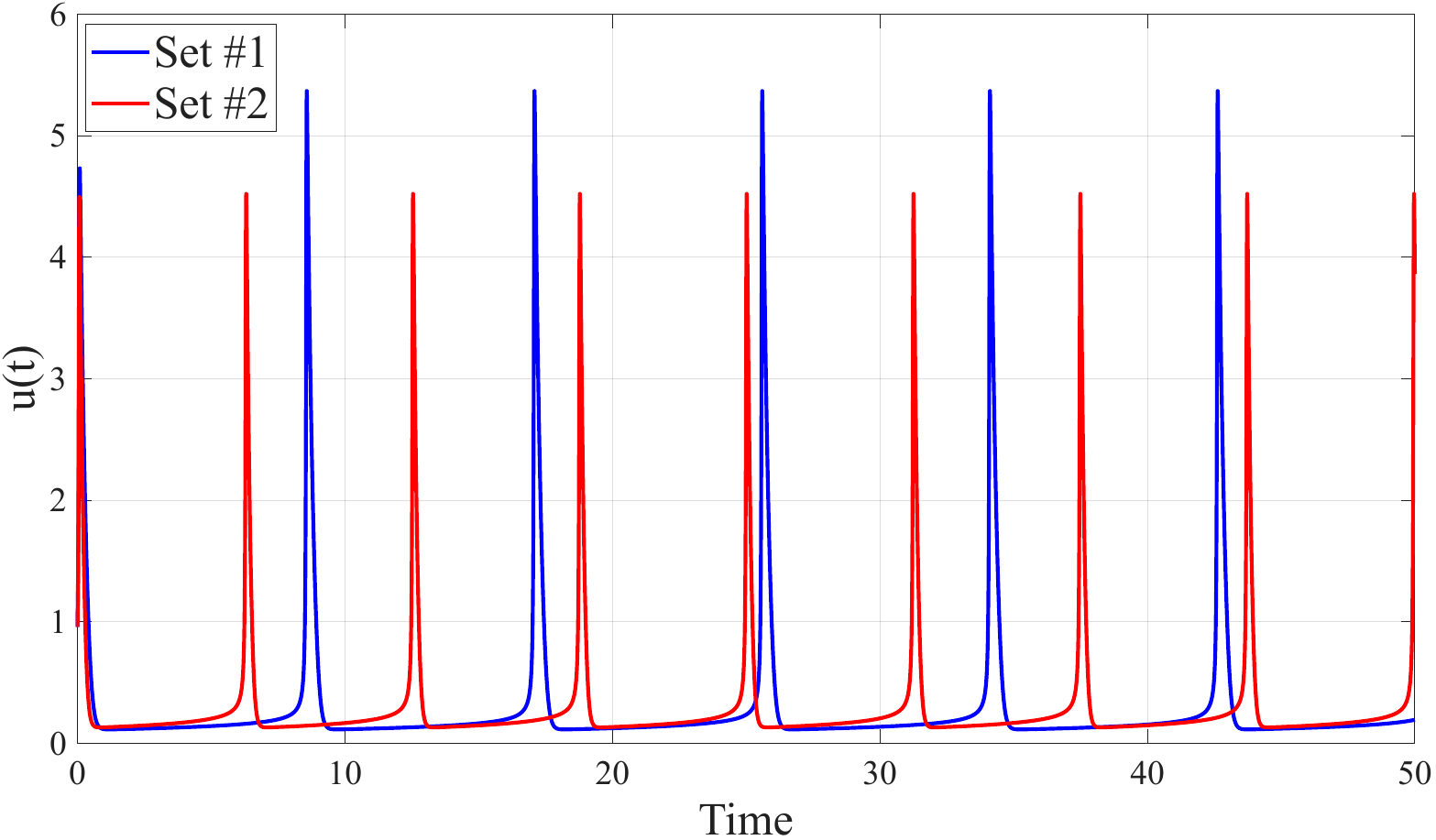}
    \caption{\textbf{Time series \(u(t)\) of two single-node Brusselator systems with identical equilibrium points but different parameter values.} The parameter set for Node 1 is \((a,b,c,d)=(1.2955,\,8.2066,\,4.9526,\,4.3203)\), and the parameter set for Node 2 is \((a,b,c,d)=(1.9522,\,9.9884,\,6.0280,\,6.5102)\).}
    \label{fig:placeholder}
\end{figure}

%%%%%%%%%%%%%%%%%%%%%%%%%%%%%%%%%%%%
%%%%%%%%%%%%%%%%%%%%%%%%%%%%%%%%%%%%

\section*{S5: Exotic patterns paramaters}

The parameter sets reported in Table~\ref{tab:model_parameters} correspond to the dynamical regimes shown in Fig.~3 of the main text. Distinct parameter tuples therefore define distinct localized node groups responsible for shaping the observed collective dynamics, including cluster synchronization, amplitude chimera, amplitude--phase chimera, and oscillon patterns.

\begin{table*}[h]
\centering
\setlength{\tabcolsep}{18pt}
\caption{Parameter sets grouped by node classes (all values rounded to 4 decimals). Distinct parameter tuples define distinct groups.}
\label{tab:model_parameters}
\begin{tabular}{lrrrrrrr}

\toprule
Set & N & $a$ & $b$ & $c$ & $d$ & $D_u$ & $D_v$ \\
\midrule

\multicolumn{8}{l}{\textbf{Cluster sync}}\\
1 & 495 & 4.4171 & 5.0403 & 0.7074 & 3.0175 & 0.5000 & 1.0000 \\
2 & 1 & 0.0764 & 3.7791 & 0.5304 & 0.0522 & 0.4811 & 1.1545 \\
3 & 1 & 0.0764 & 3.7791 & 0.5304 & 0.0522 & 0.5235 & 1.2563 \\
4 & 1 & 0.0764 & 3.7791 & 0.5304 & 0.0522 & 0.5000 & 1.2000 \\
5 & 1 & 0.0764 & 3.7791 & 0.5304 & 0.0522 & 0.8693 & 2.0863 \\
6 & 1  & 0.0764 & 3.7791 & 0.5304 & 0.0522 & 1.0606 & 2.5454 \\
\addlinespace

\multicolumn{8}{l}{\textbf{Amplitude chimera}}\\
1 & 495 & 2.8400 & 5.6603 & 2.2308 & 2.5675 & 0.5000 & 1.0000 \\
2 & 1 & 0.8259 & 4.8367 & 1.9062 & 0.7467 & 0.5000 & 1.1000 \\
3 & 1 & 1.8503 & 8.2902 & 3.2673 & 1.6727 & 0.5000 & 1.1000 \\
4 & 1 & 0.0402 & 3.9411 & 1.5533 & 0.0364 & 0.5000 & 1.1000 \\
5 & 1 & 0.7818 & 4.8592 & 1.9151 & 0.7068 & 0.5000 & 1.1000 \\
6 & 1 & 0.2508 & 4.8161 & 1.8981 & 0.2267 & 0.5000 & 1.1000 \\
\addlinespace

\multicolumn{8}{l}{\textbf{Amplitude --phase chimera}}\\
1 & 97 & 3.3681 & 3.7531 & 0.6336 & 2.2732 & 0.5000 & 1.5000 \\
2 & 1 & 2.1504 & 5.5862 & 0.9477 & 1.4513 & 0.5000 & 0.7000 \\
3 & 1 & 1.7703 & 5.0383 & 0.8547 & 1.1948 & 0.5000 & 0.7000 \\
4 & 1 & 2.2043 & 4.3802 & 0.7431 & 1.4877 & 0.5000 & 0.7000 \\
\addlinespace

\multicolumn{8}{l}{\textbf{Oscillon pattern}}\\
1 & 93 & 1.5718 & 5.3368 & 2.9803 & 5.3213 & 0.5000 & 2.0000 \\
2 & 1 & 52.6674 & 191.5065 & 106.9453 & 178.3106 & 0.5000 & 2.0000 \\
3 & 1 & 17.7655 & 66.2388 & 36.9905 & 60.1468 & 0.5000 & 2.0000 \\
4 & 1 & 50.2845 & 181.7927 & 101.5207 & 170.2429 & 0.5000 & 2.0000 \\
5 & 1 & 14.5962 & 54.8355 & 30.6225 & 49.4170 & 0.5000 & 2.0000 \\
6 & 1 & 42.9098 & 158.1963 & 88.3435 & 145.2753 & 0.5000 & 2.0000 \\
7 & 1 & 16.5602 & 62.0546 & 34.6539 & 56.0663 & 0.5000 & 2.0000 \\
8 & 1 & 1.5718 & 5.3368 & 2.9803 & 5.3213 & 0.5000 & 2.0000 \\

\end{tabular}
\end{table*}

%%%%%%%%%%%%%%%%%%%%%%%%%%%%%%%%%%%
%%%%%%%%%%%%%%%%%%%%%%%%%%%%%%%%%%%

%%%%%%%%%%%%%%%%%%%%%%%%%%%%%%%%%%%%%%%%%%
%%%%%%%%%%%%%%%%%%%%%%%%%%%%%%%%%%%%%%%%%%

\section*{S6: Empirical networks data analysis}

Table~\ref{tab:networks} summarizes the empirical networks used throughout this work together with the localization indicators introduced in Sec.~\ref{sec:measures}. For each network we report the number of nodes and edges, the inverse participation ratio (IPR), effective participation number \(N_{\mathrm{eff}}\), peak weight, tail weight, and the normalized approximation error associated with the localized spectral reduction.

\begin{table*}[h]
\centering
\scriptsize
\setlength{\tabcolsep}{2.5pt}
\caption{Summary of Networks Grouped by Domain}
\label{tab:networks}
\resizebox{\textwidth}{!}{%
\begin{tabular}{lrrrrrrrl}
\toprule
Network Name & Nodes & Edges & IPR & $N_{\mathrm{eff}}$ & Peak & Tail & NormErr & Ref \\
\midrule

\multicolumn{9}{l}{\textbf{Social}} \\
Political Blogs Network
 & 1224 & 16664 & 0.2540 & 7.0499 & 0.3981 & 0.6019 & 0.0145 & \cite{Adamic2005} \\
E-mail network for the Democratic National Convention
 & 1891 & 4465 & 0.1435 & 91.8896 & 0.2072 & 0.7928 & 0.0251 & \cite{Kunegis2016} \\
Friendship among college students in a course about leadership
 & 32 & 80 & 0.1521 & 7.5996 & 0.3054 & 0.6946 & 0.1052 & \cite{Milo2004} \\
Friendship among highschool students
 & 70 & 274 & 0.1246 & 9.9463 & 0.2611 & 0.7389 & 0.0364 & \cite{JohnsonDataRepo} \\
Online messages from an online community of students from 
 & 1899 & 13838 & 0.2205 & 11.8227 & 0.3520 & 0.6480 & 0.5517 & \cite{Opsahl2009} \\
 the University of California, Irvine\\

 \addlinespace
\multicolumn{9}{l}{\textbf{Transport}} \\
London tube network
 & 270 & 317 & 0.1066 & 19.3557 & 0.2021 & 0.7979 & 0.0348 & \cite{Williams2016} \\
Flights between airports in the United States
 & 1574 & 17215 & 0.5411 & 2.4696 & 0.6587 & 0.3413 & 0.0364 & \cite{FAA2010} \\
Paris metropolitan train grid
 & 302 & 359 & 0.1162 & 20.1148 & 0.2136 & 0.7864 & 0.5789 & \cite{Williams2016} \\
Roads Network in Rome, Italy
 & 3353 & 4831 & 0.1093 & 31.2016 & 0.2128 & 0.7872 & 0.0570 & \cite{Demetrescu2006} \\
London bike-sharing network
 & 750 & 120578 & 0.2311 & 6.7565 & 0.3829 & 0.6171 & 0.9004 & \cite{MunozMendez2018} \\
Roads Network in Sioux Falls, USA
 & 24 & 38 & 0.1490 & 6.3817 & 0.3122 & 0.6878 & 0.6752 & \cite{LeBlanc1975} \\
Roads Network in Barcelona, Spain
 & 930 & 1798 & 0.0604 & 71.3879 & 0.1270 & 0.8730 & 0.0284 & \cite{TransportationNetworks} \\
Roads Network in Winnipeg, Canada
 & 949 & 1823 & 0.4562 & 2.7037 & 0.5958 & 0.4042 & 0.0719 & \cite{TransportationNetworks} \\
Roads Network in Terrassa, Spain
 & 1603 & 2320 & 0.0438 & 80.1071 & 0.1056 & 0.8944 & 1.3457 & \cite{TransportationNetworks} \\

 \addlinespace
\multicolumn{9}{l}{\textbf{Language}} \\
Word adjacency network for Dr. Seuss's Green Eggs and Ham 
 & 50 & 93 & 0.2737 & 4.7556 & 0.4261 & 0.5739 & 0.0668 & \cite{JohnsonDataRepo} \\ book\\
 Citation network of the journal Scientometrics
 & 2729 & 10400 & 0.0749 & 42.0302 & 0.1647 & 0.8353 & 0.0492 & \cite{Schubert2002} \\
Citations to Small, Griffith and descendants
 & 1024 & 4917 & 0.1360 & 15.6264 & 0.2561 & 0.7439 & 0.0122 & \cite{Hummon1989} \\
Citations from papers that cite ”Small World Problem”
 & 233 & 994 & 0.2697 & 6.0617 & 0.4115 & 0.5885 & 0.0529 & \cite{GarfieldHistCite} \\

 \addlinespace
\multicolumn{9}{l}{\textbf{Economic}} \\
International trade network of manufactured goods
 & 24 & 195 & 0.4956 & 2.2696 & 0.6408 & 0.3592 & 0.0253 & \cite{DeNooy2018} \\
International trade network of crude animal and vegetable 
 & 24 & 195 & 0.4746 & 2.3609 & 0.6314 & 0.3686 & 0.0275 & \cite{DeNooy2018} \\  material\\
 International trade network of diplomatic exchanges
 & 24 & 199 & 0.4232 & 2.6334 & 0.5834 & 0.4166 & 0.0357 & \cite{DeNooy2018} \\
International trade network of manufactured food products
 & 24 & 201 & 0.4227 & 2.8654 & 0.5806 & 0.4194 & 0.0254 & \cite{DeNooy2018} \\
International trade network of minerals
 & 24 & 97 & 0.3662 & 3.4501 & 0.5197 & 0.4803 & 0.0642 & \cite{DeNooy2018} \\

   \addlinespace
\multicolumn{9}{l}{\textbf{Ecological}} \\
Marine Foodweb in Bahia Falsa, Mexico
 & 166 & 7707 & 0.2522 & 6.3471 & 0.3936 & 0.6064 & 0.0095 & \cite{Dunne2013} \\
Marine Foodweb in Benguela Current, South Africa
 & 29 & 198 & 0.3079 & 4.1707 & 0.4680 & 0.5320 & 0.0386 & \cite{Yodzis1998} \\
River Foodweb in Berwick Stream, New Zealand
 & 77 & 240 & 0.2642 & 5.0184 & 0.4216 & 0.5784 & 0.0418 & \cite{ThompsonTownsend2003} \\
River Foodweb in Black Rock Stream, New Zealand
 & 86 & 375 & 0.2441 & 5.1924 & 0.3933 & 0.6067 & 0.0449 & \cite{ThompsonMcintosh1998} \\
Lake Foodweb in Bridge Broom Lake
 & 25 & 106 & 0.3763 & 2.9757 & 0.5631 & 0.4369 & 0.0540 & \cite{Havens1992} \\
 River Foodweb in Broad Stream, New Zealand
 & 94 & 565 & 0.2787 & 4.7003 & 0.4244 & 0.5756 & 0.0691 & \cite{ThompsonMcintosh1998} \\
Terrestrial Foodweb in Scotch Broom, England
 & 86 & 224 & 0.3047 & 4.2344 & 0.4465 & 0.5535 & 0.0660 & \cite{Memmott2000} \\
 Fossil Assemblage Foodweb from Burgess Shale, Canada
 & 48 & 247 & 0.3234 & 3.8215 & 0.4904 & 0.5096 & 0.0505 & \cite{Dunne2008} \\
River Foodweb in Canton Creek, New Zealand
 & 102 & 697 & 0.2569 & 6.0467 & 0.3996 & 0.6004 & 0.0445 & \cite{ThompsonMcintosh1998} \\
Marine Foodweb in Carpinteria Salt Marsh Reserve, USA
 & 166 & 6557 & 0.2582 & 6.5232 & 0.4005 & 0.5995 & 0.0138 & \cite{Dunne2013} \\
River Foodweb in Caitlins Stream, New Zealand
 & 48 & 110 & 0.3422 & 3.9402 & 0.4993 & 0.5007 & 0.0619 & \cite{KlaiseJohnson2017} \\
Marine Foodweb in Cayman Islands
 & 242 & 3766 & 0.2096 & 7.4436 & 0.3528 & 0.6472 & 0.0167 & \cite{Bascompte2005} \\
Fossil Assemblage Foodweb from Chengjiang Shale, China
 & 33 & 88 & 0.2802 & 4.4061 & 0.4416 & 0.5584 & 0.0870 & \cite{Dunne2008} \\
Marine Foodweb in Chesapeake Bay, USA
 & 31 & 68 & 0.2052 & 5.1233 & 0.3879 & 0.6121 & 0.1171 & \cite{UlanowiczBaird1999} \\
River Foodweb in Coweeta, USA
 & 71 & 148 & 0.2956 & 4.8305 & 0.4295 & 0.5705 & 0.0659 & \cite{ThompsonTownsend2003} \\
 River Foodweb in Coweeta, USA
 & 58 & 126 & 0.2870 & 4.6837 & 0.4368 & 0.5632 & 0.0789 & \cite{ThompsonTownsend2003} \\
River Foodweb in Dempsters Stream during autumn, New Zealand
 & 83 & 415 & 0.2490 & 5.2384 & 0.3853 & 0.6147 & 0.0450 & \cite{KlaiseJohnson2017} \\
River Foodweb in Dempsters Stream during spring, New Zealand
 & 93 & 538 & 0.2782 & 5.2245 & 0.4191 & 0.5809 & 0.0347 & \cite{KlaiseJohnson2017} \\
River Foodweb in Dempsters Stream during summer, New Zealand
 & 107 & 966 & 0.2493 & 5.6925 & 0.4054 & 0.5946 & 0.0299 & \cite{ThompsonMcintosh1998} \\
Terrestrial Foodweb in El Verde Field Station, Puerto Rico
 & 155 & 1440 & 0.2951 & 4.9220 & 0.4421 & 0.5579 & 0.0239 & \cite{JohnsonDataRepo} \\
Marine Foodweb in Estero de Punta Banda, Mexico
 & 143 & 3412 & 0.2483 & 6.2012 & 0.3889 & 0.6111 & 0.0169 & \cite{Dunne2013} \\
Marine Foodweb in Flensburg Fjord, Germany/Denmark
 & 77 & 578 & 0.3196 & 4.3824 & 0.4631 & 0.5369 & 0.0424 & \cite{Dunne2013} \\
 Marine Foodweb in Florida Bay during dry season
 & 128 & 2106 & 0.5967 & 1.9643 & 0.7240 & 0.2760 & 0.0179 & \cite{UlanowiczDeAngelis1999} \\
River Foodweb in German Creek, New Zealand
 & 84 & 353 & 0.2378 & 5.5507 & 0.3858 & 0.6142 & 0.0477 & \cite{ThompsonMcintosh1998} \\
Terrestrial Foodweb in grasslands of the United Kingdom
 & 61 & 97 & 0.3065 & 5.2085 & 0.4404 & 0.5596 & 0.0856 & \cite{JohnsonDataRepo} \\
River Foodweb in Healy Creek, New Zealand
 & 96 & 634 & 0.2552 & 5.4068 & 0.3960 & 0.6040 & 0.0254 & \cite{ThompsonMcintosh1998} \\
 River Foodweb in Kye Burn, New Zealand
 & 98 & 629 & 0.3023 & 4.4838 & 0.4514 & 0.5486 & 0.0257 & \cite{ThompsonMcintosh1998} \\
River Foodweb in Little Kye Burn, New Zealand
 & 78 & 375 & 0.2093 & 6.1651 & 0.3423 & 0.6577 & 0.0480 & \cite{ThompsonMcintosh1998} \\
 \bottomrule
\end{tabular}%
}
\end{table*}

\addtocounter{table}{-1}

\begin{table*}[p]
\centering
\scriptsize
\setlength{\tabcolsep}{2.5pt}
\caption{Summary of Networks Grouped by Domain (continued)}
\label{tab:networks}
\resizebox{\textwidth}{!}{%
\begin{tabular}{lrrrrrrrl}
\toprule
Network Name & Nodes & Edges & IPR & $N_{\mathrm{eff}}$ & Peak & Tail & NormErr & Ref \\
\midrule

  \addlinespace
\multicolumn{9}{l}{\textbf{Ecological} \textit{(Cont.)}} \\
Lake Foodweb in Little Rock Lake, USA
 & 183 & 2452 & 0.3630 & 5.0607 & 0.4820 & 0.5180 & 0.0187 & \cite{Martinez1991} \\
Lake Foodweb in Lough Hyne, Ireland
 & 349 & 5102 & 0.2549 & 6.1405 & 0.3884 & 0.6116 & 0.0348 & \cite{Eklof2013} \\
 River Foodweb in Martins Stream, USA
 & 105 & 343 & 0.1931 & 7.2013 & 0.3240 & 0.6760 & 0.0589 & \cite{ThompsonTownsend2003} \\
 Dominance among ants
 & 16 & 36 & 0.4419 & 2.5799 & 0.6100 & 0.3900 & 0.0103 & \cite{Cole1981} \\
Dominance among bisons
 & 26 & 222 & 0.3396 & 3.4525 & 0.5025 & 0.4975 & 0.0621 & \cite{Lott1979} \\
Dominance among cattle
 & 28 & 205 & 0.3153 & 4.1664 & 0.4882 & 0.5118 & 0.0431 & \cite{ScheinFohrman1955} \\
Dominance among kangaroos
 & 17 & 91 & 0.4626 & 2.2851 & 0.6268 & 0.3732 & 0.0427 & \cite{Grant1973} \\
 Dominance among ants
 & 16 & 36 & 0.4419 & 2.5799 & 0.6100 & 0.3900 & 0.0103 & \cite{Cole1981} \\
Dominance among bisons
 & 26 & 222 & 0.3396 & 3.4525 & 0.5025 & 0.4975 & 0.0621 & \cite{Lott1979} \\
Dominance among cattle
 & 28 & 205 & 0.3153 & 4.1664 & 0.4882 & 0.5118 & 0.0431 & \cite{ScheinFohrman1955} \\
Dominance among kangaroos
 & 17 & 91 & 0.4626 & 2.2851 & 0.6268 & 0.3732 & 0.0427 & \cite{Grant1973} \\
 Dominance among macaques
 & 62 & 1167 & 0.1881 & 7.2149 & 0.3397 & 0.6603 & 0.0403 & \cite{Takahata1991} \\
Dominance among ponies
 & 17 & 129 & 0.3402 & 3.0667 & 0.5422 & 0.4578 & 0.0208 & \cite{CluttonBrock1976} \\
Dominance among sheep
 & 28 & 235 & 0.3352 & 4.0321 & 0.5004 & 0.4996 & 0.0423 & \cite{Hass1991} \\
 Dominance among wolves
 & 16 & 111 & 0.3078 & 3.6259 & 0.4864 & 0.5136 & 0.0326 & \cite{vanHooffWensing1987} \\
River Foodweb in Narrowdale Stream, New Zealand
 & 71 & 155 & 0.2439 & 6.3379 & 0.3708 & 0.6292 & 0.0522 & \cite{ThompsonTownsend2005} \\
 River Foodweb in North Col Stream, New Zealand
 & 78 & 241 & 0.3275 & 4.2632 & 0.4765 & 0.5235 & 0.0720 & \cite{ThompsonTownsend2005} \\
  Marine Foodweb in Otago Harbour, New Zealand
  & 215 & 11648 & 0.3504 & 4.4017 & 0.4906 & 0.5094 & 1.4781 & \cite{Dunne2013} \\
  River Foodweb in Powder Stream, New Zealand
 & 78 & 268 & 0.2996 & 4.0370 & 0.4630 & 0.5370 & 0.0216 & \cite{ThompsonTownsend2003} \\
Marine Foodweb in Northeast United States Shelf
 & 79 & 1396 & 0.3357 & 3.8972 & 0.4911 & 0.5089 & 0.0636 & \cite{Link2002} \\
 Lake Foodweb in Skipwith Common, England
 & 25 & 193 & 0.3052 & 3.9209 & 0.4844 & 0.5156 & 0.0337 & \cite{Warren1989} \\
Marine Foodweb in St. Marks Estuary, US
 & 48 & 221 & 0.2036 & 5.5735 & 0.3774 & 0.6226 & 0.0395 & \cite{ChristianLuczkovich1999} \\
Terrestrial Foodweb in Saint-Martin Island, Lesser Antilles
 & 42 & 205 & 0.2846 & 5.9639 & 0.4331 & 0.5669 & 0.0480 & \cite{GoldwasserRoughgarden1993} \\
River Foodweb in Stony Stream, New Zealand
 & 109 & 829 & 0.2974 & 4.7504 & 0.4472 & 0.5528 & 0.1687 & \cite{ThompsonMcintosh1998} \\
River Foodweb in Stony Stream, New Zealand
 & 112 & 832 & 0.2958 & 4.7437 & 0.4439 & 0.5561 & 1.4906 & \cite{ThompsonMcintosh1998} \\
River Foodweb in Sutton Stream during autumn, New Zealand
 & 80 & 335 & 0.2702 & 4.5114 & 0.4183 & 0.5817 & 0.0157 & \cite{KlaiseJohnson2017} \\
River Foodweb in Sutton Stream during spring, New Zealand
 & 74 & 391 & 0.3133 & 4.5380 & 0.4531 & 0.5469 & 0.0284 & \cite{KlaiseJohnson2017} \\
River Foodweb in Sutton Stream during summer, New Zealand
 & 87 & 424 & 0.3621 & 3.7374 & 0.4810 & 0.5190 & 0.0173 & \cite{ThompsonMcintosh1998} \\
Marine Foodweb in Sylt Tidal Basin, Germany
 & 215 & 11567 & 0.3278 & 4.7449 & 0.4640 & 0.5360 & 0.0529 & \cite{Dunne2013} \\
River Foodweb in Troy Stream, USA
 & 77 & 181 & 0.2245 & 5.8638 & 0.3600 & 0.6400 & 0.0464 & \cite{ThompsonTownsend2003} \\
River Foodweb in Venlaw Stream, New Zealand
 & 66 & 187 & 0.2652 & 4.8426 & 0.4269 & 0.5731 & 0.0297 & \cite{ThompsonTownsend2003} \\
Marine Foodweb in Weddel Sea, Antarctica
 & 483 & 15062 & 0.3009 & 5.2063 & 0.4259 & 0.5741 & 0.0356 & \cite{Eklof2013} \\
Marine Foodweb in Ythan Estuary, Scotland
 & 82 & 394 & 0.2807 & 4.6295 & 0.4340 & 0.5660 & 0.0298 & \cite{Huxham1996} \\
Marine Foodweb in Ythan Estuary, Scotland
 & 166 & 7275 & 0.2665 & 5.7474 & 0.4138 & 0.5862 & 0.0511 & \cite{Dunne2013} \\

   \addlinespace
\multicolumn{9}{l}{\textbf{Biological}} \\
Human protein-protein interactome produced
\\ by a mass spectrometry-based approach by Figeys et al.
 & 2239 & 6432 & 0.1045 & 74.4869 & 0.1891 & 0.8109 & 0.0090 & \cite{Ewing2007} \\
Neuronal network for a mouse brain
 & 213 & 16242 & 0.1940 & 7.5993 & 0.3398 & 0.6602 & 0.0365 & \cite{JohnsonDataRepo} \\
Human gene regulatory network for a person with cancer
 & 4049 & 11706 & 0.0477 & 202.2597 & 0.0952 & 0.9048 & 0.0693 & \cite{Gerstein2012} \\
Neuronal network for Caenorhabditis elegans
 & 297 & 2148 & 0.1405 & 14.6583 & 0.2581 & 0.7419 & 0.0090 & \cite{WattsStrogatz1998} \\
Gene regulatory network for Escherichia coli from Thieffry et al.
 & 418 & 519 & 0.1599 & 19.4156 & 0.2774 & 0.7226 & 0.0092 & \cite{Thieffry1998} \\
Gene regulatory network for Escherichia coli from Salgado et al.
 & 1470 & 2902 & 0.1134 & 62.0735 & 0.1832 & 0.8168 & 0.0117 & \cite{Salgado2013} \\
Gene regulatory network for Mycobacterium tuberculosis
 & 1624 & 3163 & 0.0712 & 99.6479 & 0.1260 & 0.8740 & 0.0187 & \cite{Sanz2011} \\
Human gene regulatory network for a healthy person
 & 4071 & 8465 & 0.0375 & 240.1826 & 0.0813 & 0.9187 & 0.0251 & \cite{Gerstein2012} \\
Gene regulatory network for Pseudomonas aeruginosa
 & 691 & 984 & 0.1286 & 45.9039 & 0.2089 & 0.7911 & 0.0731 & \cite{GalanVasquez2011} \\
Gene regulatory network for Saccharomyces cerevisiae from Harbison et al. & 2933 & 6150 & 0.0696 & 152.1040 & 0.1220 & 0.8780 & 0.0184 & \cite{Harbison2004} \\
Metabolic network for Saccharomyces cerevisiae
 & 1510 & 3807 & 0.0883 & 46.8165 & 0.1663 & 0.8337 & 0.0086 & \cite{Jeong2000} \\
Gene regulatory network for Saccharomyces cerevisiae from Constanzo et al.
 & 688 & 1078 & 0.1360 & 32.8261 & 0.2307 & 0.7693 & 0.0086 & \cite{Costanzo2001} \\
Connectome of the Rhesus brain, extracted from tract
\\ tracing studies collated in the CoCoMac database
 & 242 & 3054 & 0.2252 & 7.4285 & 0.3662 & 0.6338 & 0.0154 & \cite{Harriger2012} \\
Connectome of the Rhesus brain, via a retrograde tracer study
 & 91 & 582 & 0.3490 & 4.0950 & 0.4772 & 0.5228 & 0.0365 & \cite{Markov2013} \\
Mouse’s primary visual cortex connectome 1
 & 503 & 23030 & 0.3783 & 4.2415 & 0.5085 & 0.4915 & 0.0691 & \cite{BotaSwanson2007} \\
Mouse’s primary visual cortex connectome 2
 & 502 & 24656 & 0.3764 & 4.0911 & 0.5090 & 0.4910 & 0.0095 & \cite{Carere2007} \\
Mouse’s primary visual cortex connectome 3
 & 493 & 25988 & 0.3782 & 4.0854 & 0.5084 & 0.4916 & 0.0107 & \cite{Cowell2007} \\
\bottomrule
\end{tabular}
}
\end{table*}

\FloatBarrier

\bibliographystyle{apsrev4-2}
\bibliography{references}

\end{document}